\documentclass[a4paper,12pt]{article}
\usepackage{jheppub,booktabs}
\usepackage{amsmath,amssymb}
\usepackage{xcolor,colortbl}
\usepackage{cancel}
\usepackage{subfigure}
\usepackage{mathrsfs}
\usepackage{graphicx}
\usepackage{multicol}
\usepackage{multirow}
\usepackage{pdflscape}
\usepackage{hepunits}
\usepackage{slashed}
\usepackage{feynmf}
\usepackage{enumerate}
\usepackage{siunitx}
\usepackage{physics}
\usepackage{bbm}
\usepackage{appendix}
\usepackage[utf8x]{inputenc}
\usepackage{subfigure}
\definecolor{myblue}{RGB}{0,76,153}
\definecolor{myred}{RGB}{166,41,41}
\usepackage{hyperref}
\hypersetup{colorlinks=true, citecolor=myred, linkcolor=myblue, urlcolor=myblue}

\DeclareMathOperator{\diag}{diag}

\newcount\colveccount
\newcommand*\colvec[1]{
	\global\colveccount#1
	\begin{pmatrix}
		\colvecnext
	}
	\def\colvecnext#1{
		#1
		\global\advance\colveccount-1
		\ifnum\colveccount>0
		\\
		\expandafter\colvecnext
		\else
	\end{pmatrix}
	\fi
}
\newcommand*\xbar[1]{%
	~
	\hbox{%
		\vbox{%
			\hrule height 0.5pt 
			\kern0.3ex
			\hbox{%
				\kern-0.1em
				\ensuremath{#1}%
				\kern-0.1em
			}%
		}%
	}%
	~
} 

\newcommand{\sL}{s_{L}}
\newcommand{\sLbar}{\bar{s}_{L}}
\newcommand{\chibar}{\bar{\chi}}
\newcommand{\chin}{\chi_n}

\newcommand{\psibar}{\bar{\psi}}

\newcommand{\psibarn}{\bar{\psi}_n}
\newcommand{\sR}{s_{R}}
\newcommand{\sRbar}{\bar{s}_{R}}
\newcommand{\sigmabar}{\bar{\sigma}}
\newcommand{\gni}{g_n^i}
\newcommand{\tLni}{t_{\text{L},n}^i}
\newcommand{\tRni}{t_{\text{R},n}^i}

\newcommand{\fni}{f_n^i}
\newcommand{\gn}{g_n}
\newcommand{\tLn}{t_{\text{L},n}}
\newcommand{\tLntilde}{\tilde{t}_{\text{L},n}}
\newcommand{\tRn}{t_{\text{R},n}}
\newcommand{\fn}{f_n}

\newcommand{\evalat}[2]{#1_{|_{#2}}}
\newcommand{\evalsub}[2]{#1\vphantom{{}}_{|_{#2}}}
\newcommand{\shouldeq}{\overset{!}{=}}

\newcommand{\BesselJ}[2]{J_{#1}(#2)}
\newcommand{\BesselY}[2]{Y_{#1}(#2)}
\newcommand{\BesselI}[2]{I_{#1}(#2)}
\newcommand{\BesselK}[2]{K_{#1}(#2)}

\author[a]{Simone Blasi,}
\author[b,c]{Julian Bollig,}
\author[c]{and Florian Goertz}
\affiliation[a]{Theoretische Natuurkunde and IIHE/ELEM, Vrije Universiteit Brussel, \& The  International Solvay Institutes, Pleinlaan 2, B-1050 Brussels, Belgium}
\affiliation[b]{Physikalisches Institut, Albert-Ludwigs-Universit{\"a}t Freiburg, Hermann-Herder-Straße 3, 79014 Freiburg}
\affiliation[c]{Max-Planck-Institut f{\"u}r Kernphysik, Saupfercheckweg 1, 69117 Heidelberg, Germany}
\emailAdd{simone.blasi@vub.be}\emailAdd{julian.bollig@physik.uni-freiburg.de}
\emailAdd{florian.goertz@mpi-hd.mpg.de}

\title{Holographic Composite Higgs Model Building: Soft Breaking, Maximal Symmetry, and the Higgs Mass}

\abstract{We study the emergence and phenomenological consequences of recently proposed new structures, namely soft breaking of the Higgs shift symmetry and `maximal symmetry' of the composite sector, in holographic realizations of composite Higgs models.
For the former, we show that soft breaking can also successfully be implemented in a full 5D warped model, where symmetry-restoring universal boundary conditions for the fermion fields allow to break the problematic connection between a realistically light Higgs and anomalously light top partners.
For the latter, we demonstrate that the minimal incarnation of maximal symmetry in the holographic dual leads to a sharp prediction of $m_h\approx 197$ GeV for $f=800\,$GeV. We find that a viable implementation is possible with sizable negative gauge brane kinetic terms, allowing for $m_h=125$\,GeV.
Overall, both approaches offer promising directions to improve the naturalness also of holographic realizations of composite Higgs models.}

\date{\today}

\begin{document}
\maketitle


\section{Introduction}
 Explaining the lightness of the Higgs scalar~\cite{ATLAS:2012yve,CMS:2012qbp} by assuming it to be a composite pseudo-Nambu-Goldstone boson of a global symmetry~$G$ spontaneously broken to a subgroup~$H$ by condensation in a strongly coupled sector \cite{Kaplan:1983fs,Kaplan:1983sm,Dugan:1984hq}, is an attractive path already realized in nature in the context of the light pions of QCD. 
 The Higgs mass is thus saturated in the infra-red (IR) and its Goldstone nature addresses the mass gap to other yet undiscovered resonances. 
 
 However, concrete composite Higgs models (CHM) face challenges in generating a fully viable Higgs potential, which here becomes calculable, being induced from interactions with the Standard Model (SM) particles explicitly breaking the global symmetry $G$. In particular, the linear mixings of the top quark with the fermionic operators of the strongly coupled sector, following the partial-compositeness paradigm \cite{Kaplan:1991dc,Agashe:2004rs,Contino:2003ve,Contino:2006qr}, induce significant explicit breaking of $G$ due to the missing $G-$partners of the top quark, which generically makes the Higgs boson too heavy. 
 This holds especially in variants of the so-called minimal CHM (MCHM)~\cite{Agashe:2004rs}, where $G=SO(5)$ and $H=SO(4)$ (which is the simplest setup featuring custodial symmetry), on which we will focus in the following.\footnote{For a review of CHMs and a summary of the most important constraints, we direct the reader towards \cite{Bellazzini2014,Panico:2015jxa,Goertz:2018dyw} and the references therein.}  
 To reduce the Higgs mass to a realistic value requires then rather light top partners, significantly below the generic resonance masses, in order to reduce the explicit breaking while still reproducing the large top-quark mass~\cite{Contino:2006qr,Matsedonskyi:2012ym,Pomarol:2012qf,Csaki:2008zd,DeCurtis:2011yx,Carmona2015} (see, however, \cite{Bally:2022naz}). 
 
 Because those partners remain elusive at the LHC so far, with the latest bounds already surpassing $m_T\gtrsim 1.5$\,TeV \cite{CMS:2022fck,ATLAS:2022ozf,ATLAS:2022hnn}, the global symmetry breaking scale $f$ is raised similarly beyond a TeV. 
 This puts CHMs under increasing pressure since the tuning scales as $\xi\equiv v^2/f^2$, with $v=246\,$GeV the electroweak vacuum expectation value (vev), and is pushed towards the per cent level. Beyond that, it prevents a potential exploration of the spectrum of resonances, residing above the top partners, at the LHC in the minimal scenario.
 Another issue of minimal models, explored in this article, is their enhanced tuning compared to the minimal estimate of $\xi$, rooted in the fact that a non-trivial vev arises only from an intricate interplay of actually hierarchically different terms in the potential (see below).
 
In recent works \cite{Csaki2017,Csaki2018,Blasi2019,Blasi2020}, both these problems have been addressed in effective 4D models, on the one hand by taming the explicit global symmetry breaking, turning its source into a soft mass term in the Lagrangian \cite{Blasi2019,Blasi2020} and thereby removing the stringent connection between the top mass and the size of the Higgs mass.
This was achieved by augmenting the top-quark embedding into a full representation of $SO(5)$, where the undiscovered $G$-partners were lifted above the TeV scale via soft mass terms, instead of fully decoupling them~\cite{Blasi2019,Blasi2020}.
On the other hand, constructing models which feature an enhanced `maximal symmetry' of the composite sector, removes the mismatch of terms responsible for generating the electroweak vev \cite{Csaki2017,Csaki2018}, as explained below.
Combining these concepts, as put forward in \cite{Blasi2020}, allows to keep the scale of the model below a TeV with minimal (${\cal O}(10\%)$) tuning and thus to still have potential rich discoveries around the corner. 

So far, all these works focused on 4D implementations of the CHMs, based on low-energy effective descriptions. In this article we will provide the first full holographic 5D implementations of these ideas, delivering a calculable dual description of the strongly coupled theories, following the AdS/CFT correspondence \cite{Maldacena:1997re,Gubser:1998bc,Witten:1998qj}. In such dual 5D models, the Higgs boson is realized as a scalar component of a 5D gauge field in a compactified extra dimension~\cite{Contino:2010rs,Contino:2003ve,Agashe:2004rs}, as in models of gauge-Higgs unification~\cite{Manton:1979kb,Fairlie:1979at,Hosotani:1983vn,Hosotani:1983xw}, with the bulk gauge symmetry $G$ broken to $H$ by boundary conditions (BCs) at the IR boundary of the theory. In the holographic dictionary, this boundary as well as the 5D bulk of the space-time correspond to the strongly coupled CFT (spontaneously broken by the IR cutoff), i.e. the composite sector. The resonances of this strong sector are dual to Kaluza-Klein (KK) excitations in the compactified extra dimension, residing close to the IR boundary and the higher dimensional gauge symmetry is dual to the protecting Goldstone shift-symmetry.
The ultra-violet (UV) boundary on the other hand, corresponds to the elementary sector where the light SM particles are mostly located, while they mix with the composite resonances by leaking into the bulk. The light SM-like states thus correspond to zero modes of 5D bulk fields.

A von Neumann (Dirichlet) BC for a gauge field, denoted by a $+$ $(-)$ in the following, on the UV boundary corresponds to the symmetry being (not) gauged, thus at least the gauge degrees of freedom corresponding to the SM group need to feature + BCs (which is minimally realized in the MCHM). Moreover a $-$ BC on the IR boundary corresponds to the respective symmetry being spontaneously broken. Massless fields, i.e. zero modes, just appear for $[++]$ BCs. In the MCHM, the latter are employed exactly for the SM spectrum of gauge bosons and fermions. The $G-$partners of the top quark, filling the bulk $SO(5)$ representations, accordingly feature mixed $[-+]$ or $[+-]$ BCs at the end of the extra dimension, which is exactly the $G-$breaking source of the pNGB Higgs potential in the holographic dual.

As mentioned, the works \cite{Blasi2019,Blasi2020} pointed out that it is possible to soften this breaking by restoring the $G-$partners of the top in the spectrum of elementary fields, which was shown to allow for a lighter Higgs boson for given top-partner mass. In the holographic dual, this symmetry restoration is realized by assigning universal boundary conditions to the fermionic $G-$representations and lifting the set of additional zero modes via UV-brane soft mass terms \cite{Blasi2020}, which we will explore in detail below. Performing full scans of the parameter space, we will provide numerical results for the top-partner spectrum and the Higgs mass as well as the resulting tuning in the 5D realization of the composite Higgs, demonstrating how the universal BCs allow to bring down the Higgs mass.

We will also discuss how maximal symmetry emerges in 5D models from a relation between (composite) IR-brane masses and provide a tight prediction for the mass of the Higgs boson that results from the restricted structure. We will in particular discuss possibilities to shift this, via gauge brane kinetic terms (BKTs), while still benefiting from the reduced tuning due to maximal symmetry. Finally, we will demonstrate the virtues of combining universal BCs with maximal symmetry for realizing a viable electroweak symmetry breaking (EWSB).

This article is organized as follows. In Section~\ref{sec:softbreaking}, after recapitulating the basics of holographic CHMs and soft breaking, we introduce the soft version of the MCHM (the sMCHM) in 5D and present its phenomenology, as anticipated above. We also study the potential emergence of ghost/tachyonic states due to negative BKTs. We then turn to the 5D structure of maximal symmetry and its phenomenological consequences in Section~\ref{sec:maxsym}, including its soft-breaking version. We finally conclude in Section~\ref{sec:conc}. A series of appendices summarizes our (spinor) conventions, the solutions for the 5D wave functions, as well as the numerical approach for calculating the Higgs potential.


\section{Soft breaking in holographic Composite Higgs Models}
\label{sec:softbreaking}

To study the impact of softened symmetry breaking (and later on also maximal symmetry) on CHMs which suffer from too light top partners and a high tuning, we stick in the following to the MCHM, considering in particular the MCHM$_5$~\cite{Contino:2006qr} as our starting point, where the subscript $5$ denotes that we insert fermions as 5-plets under the global $SO(5)$ symmetry of the model. Since this is a rather minimal incarnation of the composite Higgs idea, it is in fact an ideal candidate for the investigation of new features which could be carried over to more complex models in future work. 


\subsection{Holographic realization of softly broken CHMs}

\noindent
For the realization of an MCHM$_5$ which features softened symmetry breaking in 5D (from now on called sMCHM$_5$), we consider the usual, conformally flat, truncated AdS$_5$ space
\begin{equation}
	\dd s^2= a(z)^2(\eta_{\mu\nu}\dd x^\mu\dd x^\nu-\dd z^2)\,,
	\label{eq:confmetric}
\end{equation}
with $a(z)=R/z$, $z\in[R,R']$ being the coordinate of the compact dimension and $R$ ($R'$) indicating the position of the UV (IR) brane.\footnote{See, e.g.,  \cite{Contino:2010rs,Contino:2003ve,Agashe:2004rs,Contino:2004vy,Falkowski:2006vi,Falkowski:2007kd,Medina:2007hz,Panico:2007qd,Gherghetta:2010cj} for more details on the holographic implementation of CHMs.} We fix the UV scale as the inverse Plank scale, $R=1/M_{Pl}\sim \SI{e-16}{\TeV^{-1}}$, and keep in the following $R'\sim\SI{1}{\TeV^{-1}}$, such as to solve the hierarchy problem. The latter is related to the $SO(5)$ symmetry breaking scale as (see e.g. \cite{Falkowski:2006vi})
\begin{equation}
    \label{eq:fRprimeconnection}
    f=\frac{2\sqrt{R}}{g_5\sqrt{R'^2-R^2}}\approx \frac{2}{g_*R'}\,,
\end{equation}
where $g_5$ is the dimensionfull and $g_*=g_5/\sqrt{R}$ the dimensionless coupling constant of $SO(5)$. In the following we will be interested in breaking scales of $f\sim \SI{800}{\GeV}$, which corresponds to a vacuum misalignment of $\xi\equiv v^2/f^2\sim 0.1$ in the dual 4D model, in agreement with bounds on electroweak precision operators (see eg. \cite{Grojean2013}).\par

Further we choose \emph{universal} BCs (see \cite{Csaki2004}) for the fermionic $SO(5)$ multiplets in fundamental $\mathbf{5}$ representations\footnote{Here, the generators $\tilde{T}^A$, under which the multiplets transform, have been converted into a diagonal basis (see for example Appendix of \cite{Carmona2015}). Moreover, here and in the following, we suppress the additional $U(1)_X$ group required to realize the correct SM hypercharges.}
\begin{equation}
		\psi_L= \colvec{3}{w_L[+,+]}{q_L[+,+]}{s_L[+,+]}\qc \psi_R= \colvec{3}{w_R[+,+]}{v_R[+,+]}{t_R[+,+]},
		\label{eq:SO5fields}
\end{equation}
where signs in square brackets denote the BCs on the [UV,IR]-brane,  
which avoids explicit hard breaking of the $SO(5)$ symmetry \cite{Blasi2019}. Note that there are also the opposite chirality fields, due to fermions in 5D being vector-like, which feature opposite BCs. The 5D fields can be related with the fields of a 4D CHM by performing a  KK decomposition. The zero modes of the $SU(2)_L$ doublet $q_L=(t_L,b_L)^T$ and singlet $t_R$ then correspond to the eponymous 4D SM fields. They will obtain their masses due to interactions of the fermionic multiplets with the Higgs field on the IR-brane,  while the additional zero modes appearing for $[+,+]$ fields will be lifted to the TeV scale due to \emph{soft boundary-mass terms} \cite{Blasi2019}. This will allow to substantially mitigate the contribution of the top sector to the Higgs boson mass (see below). 

Note that in principle we could also introduce similar $SO(5)$ fields for the other flavor states in the SM. However, since $y_t$ is much bigger than any other Yukawa coupling in the SM, making the couplings of the top quark to the composite Higgs most important, we will restrict ourselves to this sector and treat other contributions as small corrections which are beyond the scope of this work. The zero-modes of the $v$ and $w$ doublets as well as the $s$ singlet correspond to new elementary fermions in the 4D theory, see \cite{Blasi2019} (as opposed to the higher KK excitations, which correspond to composite resonances). Unlike the SM fermions, they have not been observed in nature, yet, which means that their masses have to be uplifted from a model building perspective. In the sMCHM$_5$, this is achieved through couplings with 4D localized fields on the UV-brane, which correspond to vector-like mass terms and can be translated into dynamical boundary conditions for the 5D fields as we will explain later. In the limit where these mass terms approach infinity, the model converges to the original MCHM$_5$ while in the case they vanish, the global symmetry gets fully restored such that the Higgs becomes massless (ignoring gauge effects). Finite values correspond to an intermediate situation with reduced Higgs mass induced via the soft-breaking masses, taming the UV behavior of the Higgs potential, see Ref.~\cite{Blasi2019}.


\subsection{The mass of the lightest top partner in holographic models}

We first strive to find the mass of the lightest top partner in the holographic sMCHM$_5$ as a prime object of concern. To achieve this, we take a look at the relevant part of the action away from the boundaries, which reads
\begin{equation}
	\label{eq:warpedbulkaction}
	S_\text{bulk}\!=\!\int \!\dd[5]x\sqrt{G}\qty\bigg{\sum_{k=1,2} \frac{i}{2}E^M_A(\bar{\Psi}_k\Gamma^AD_M\Psi_k-D_M\bar{\Psi}_k\Gamma^A\Psi_k)-R (c_\psi\bar{\Psi}_1\Psi_1 + c_\chi \bar{\Psi}_2\Psi_2)},
\end{equation}
where capital letters denote 5D indices, $\sqrt{G}=(\det g_{MN})^{1/2}=a(z)^5$ is the square-root of the metric determinant, $g_{MN}=a(z)^2\eta_{MN}$ is the conformal 5D metric defined by Eq. \eqref{eq:confmetric} and $E_A^M=a^{-1}(z)\delta_A^M$ is the inverse 5D vielbein given by $E_M^A\eta_{AB}E_N^B=g_{MN}$.
Moreover, the two $SO(5)$ multiplets introduced before are embedded into separate 5D spinors 
\begin{equation}
	\label{eq:5Dspinors}
	\Psi_1 = \colvec{2}{\psi_L}{\psibar}\qc \Psi_2 = \colvec{2}{\chi}{\psibar_R}
\end{equation}
in supersymmetric notation, where we have omitted the indices on the fields (we follow the conventions of \cite{Csaki2004}: see Appendix \ref{sec:appDiracMatrices} for the contraction rules as well as the definition of the 5D gamma matrices $\Gamma^A$). The dimensionless bulk masses are denoted by $c_{\psi,\chi}$, where we chose $c_{\psi},c_{\chi}\in (-1/2,1/2)$, such that all fields reside close to the IR-brane. Because we are focusing in the following on fermionic contributions, the gauge fields will be ignored for now, i.e. $D_M\to\partial_M$.

By explicitly splitting the 5D fields into their KK-modes as
\begin{alignat}{2}
	\psi_L^i &=\sum_n \tLni(z)\chin^i(x)\qquad \psibar^i &&=\sum_n \fni(z)\psibarn^i(x)\\
	\chi^i&=\sum_n \gni(z)\chin^i(x)\qquad \psibar_R^i&&=\sum_n \tRni(z)\psibarn^i(x)\,,
\end{alignat}
where $i=1,...,5$ is an $SO(5)$ index, we can derive the equations of motion for the $z$-dependent parts by variation of the bulk action 
\begin{alignat}{3}
		\tLn'+\frac{c_\psi-2}{z} \tLn-m_n\fn &=0&&\qquad  \fn'-\frac{c_\psi+2}{z} \fn+m_n\tLn&&=0\label{eq:warpedcasemixedEOM1}\\
		\gn'+\frac{c_\chi-2}{z} \gn-m_n\tRn &=0 &&\qquad \tRn'-\frac{c_\chi+2}{z} \tRn+m_n\gn &&=0\label{eq:warpedcasemixedEOM2}\,,
	\end{alignat}
where the KK-modes $\{\chin\}$ and $\{\psibarn\}$ provide a chiral basis for left- and right-handed 4D fields which obey the Dirac equations
\begin{align}
	-i\sigmabar^\mu\partial_\mu\chin+m_n\psibarn &=0 \\
		-i\sigma^\mu\partial_\mu\psibarn+m_n\chin &=0.
\end{align}
The $m_n$ denote the masses of the KK-tower for the fermionic fields. Note that we suppressed the $SO(5)$ index in the latter equations for clarity. We finally decouple the mixed equations of motion in Eqs. \eqref{eq:warpedcasemixedEOM1} and \eqref{eq:warpedcasemixedEOM2}, leading to
\begin{align}
	\tLn''-\frac{4}{z}\tLn'+\qty\bigg(m_n^2-\frac{c_\psi^2+c_\psi-6}{z^2})\tLn &=0 \\ \fn''-\frac{4}{z}\fn'+\qty\bigg(m_n^2-\frac{c_\psi^2-c_\psi-6}{z^2})\fn &=0\,,\label{eq:warpedcasedecoupledEOM}
\end{align}
and equivalently for $\gn$ and $\tRn$. 

The equations above are Bessel equations such that the general solutions for the $z$-dependent fields can be expressed in terms of Bessel functions of $1^\text{st}$ and $2^\text{nd}$ kind
\begin{align}
		\label{eq:EOMsolutionwarpedtLn}
		\tLni(z)&=\qty\bigg(\frac{R}{z})^{{c_\psi}-2}\qty\Big(A_n^iC_{c_\psi}(z)+B_n^iS_{c_\psi}(z))\\
		\label{eq:EOMsolutionwarpedfn}
		\fni(z)&=\qty\bigg(\frac{R}{z})^{-c_\psi-2}\qty\Big(C_n^iC_{-c_{\psi}}(z)+D_n^iS_{-c_{\psi}}(z))\\
		\label{eq:EOMsolutionwarpedgn}
		\gni(z)&=\qty\bigg(\frac{R}{z})^{c_\chi-2}\qty\Big(F_n^iC_{c_\chi}(z)+G_n^iS_{c_\chi}(z))\\
		\label{eq:EOMsolutionwarpedtRn}
		\tRni(z)&=\qty\bigg(\frac{R}{z})^{-c_\chi-2}\qty\Big(H_n^iC_{-c_\chi}(z)+I_n^iS_{-c_\chi}(z)),
	\end{align}
where we used warped trigonometric functions $S_c(z)$ and $C_c(z)$ as a basis because of their useful properties $S_c(R)=0$, $C_c(R)=1$, $S_c'(R)=m_n$ and $C_c'(R)=0$ (see their definition in Appendix \ref{sec:appwarpedsinecosine}). 
We can, therefore, immediately see that we can reduce our degrees of freedom (dofs) drastically by evaluation of Eqs. \eqref{eq:warpedcasemixedEOM1} and \eqref{eq:warpedcasemixedEOM2} at $z=R$, which relates the coefficients as
\begin{equation}
	\label{eq:warpedmixedrelations}
	A_n^i=-D_n^i \qquad B_n^i=C_n^i  \qquad F_n^i=-I_n^i \qquad G_n^i=H_n^i.
\end{equation}\par
So far we have neglected all gauge terms in the covariant derivative. Including the contributions from the broken generators $D_5=\partial_z-ig_5 \tilde{T}^{\hat{a}} C^{\hat{a}}_5(x,z)$, $\hat{a}=1,...,4$, introduces a Yukawa term to the action,  where $\tilde{T}^{\hat{a}}$ are the broken generators of $SO(5)/SO(4)$. The gauge field zero modes $C^{\hat{a}(0)}_5(x,z)\equiv f_h^{\hat{a}}(z)h^{\hat{a}}(x)$ incorporate the four dofs of the Higgs field $h^{\hat{a}}(x)$ with bulk profiles $f^{\hat{a}}_h(z)$ and vev $\expval{ h^{\hat{a}}(x)}=\tilde{v}\delta^{\hat{a}4}$. These additional interactions mix the different components of the in general $h$-dependent 5D Dirac fields $\psi_{L,R}$ of Eq.~\eqref{eq:SO5fields} in the bulk making it in general extremely hard to solve the bulk equations of motion. 

Fortunately, by performing a 5D gauge transformation \cite{Carmona2015}
\begin{equation}
    C^{\hat{a}}_M(x,z)\to \Omega(z) C^{\hat{a}}_M(x,z)\Omega(z)^T-\frac{i}{g_5}(\partial_M\Omega(z))\Omega(z)^T
\end{equation}
with 
\begin{equation}
	\Omega(z)=\exp(-ig_5\expval{ h^{\hat{a}}(x)}\tilde{T}^{\hat{a}}\int_R^z\dd{z'}f^{\hat{a}}_h(z^\prime))=\exp(-i\frac{\tilde{v}}{f(z)}\sqrt{2}\tilde{T}^{\hat 4})
\end{equation}
and equivalently for all other gauge fields present, we are able to remove the Higgs vev in the bulk due to 5D gauge invariance (see e.g. \cite{Falkowski:2006vi} for a more detailed explanation). The function $f(z)$ is defined as
\begin{equation}
    f(z)\equiv \left[\frac{g_5}{\sqrt{2}}\int_R^{z}\dd{z'}f^{\hat 4}_h(z')\right]^{-1}
\end{equation}
where the bulk profile of the Higgs field yields
\begin{equation}
    f^{\hat{4}}_h(z)=\frac{z}{R}\left[\int_R^{R'} \dd{z'} \frac{z'}{R}\right]^{-1/2}\,,
\end{equation}
and reduces to the $SO(5)$ symmetry breaking scale defined in Eq. \eqref{eq:fRprimeconnection} on the IR brane (i.e. $f=f(R')$).  
However, this gauge transformation will introduce mixing of fermionic states on the IR brane due to the corresponding Wilson line transformation $\evalsub{F}{z=R'}\to W_{\text{IR}}^\dagger\evalsub{F}{z=R'}$ of the fermionic 5D fields with $W_{\text{IR}}\equiv \Omega(R')$ as we will see further below. 

The action on the IR brane is given by
\begin{align}
\label{eq:SIR}
	S_\text{IR} = &\int \dd[4]{x}\sqrt{-g_\text{IR}}\left[\bar{\Psi}_1
	\begin{pmatrix}
		C_{12} & 0 \\
		0 & C_{12}
	\end{pmatrix}
	\Psi_2+\bar{\Psi}_2
	\begin{pmatrix}
		C_{12} & 0 \\
		0 & C_{12}
	\end{pmatrix}
	\Psi_1\right.\nonumber\\
	&\hphantom{\int \dd[4]{x}\sqrt{-g_\text{IR}}\left[\right.}\left.+iR'\bar{\Psi}_1\gamma^{\mu}
	\begin{pmatrix}
		\kappa_L\mathbbm{1}_5 & 0 \\
		0 & 0
	\end{pmatrix}
	\partial_\mu\Psi_1+iR'\bar{\Psi}_2\gamma^{\mu}
	\begin{pmatrix}
		0 & 0 \\
		0 & \kappa_R\mathbbm{1}_5
	\end{pmatrix}
	\partial_\mu\Psi_2\right]_{z=R'},
\end{align}
with $\sqrt{-g_\text{IR}}=(R/R')^4$ the 4D metric determinant. The diagonal matrix $C_{12}=\diag(c_1,c_1,c_1,$ $c_1,c_2)$ parametrizes the boundary mass terms between the spinors of Eq.~\eqref{eq:SO5fields}, as required for a massive SM-like fermion spectrum.\footnote{Note that the structure of $C_{12}$ is chosen to preserve $H=SO(4)\cross U(1)_X$ symmetry in the fermionic Lagrangian. See e.g. \cite{Carmona2015} for further details.} We further demand that $c_1\neq c_2$ (which would restore the full $SO(5)$ symmetry otherwise) in order to generate a proper Higgs potential.
Note that we have also included fermionic BKTs for the left- and right-handed fields, parametrized by $\kappa_L, \kappa_R$, to have the additional possibility to modify the masses of the non-zero modes in the fermionic KK-towers\footnote{The $\kappa_L,\kappa_R$ correspond in the 4D dual to a change of $g_\Psi$, which defines the mass scale of the fermionic resonances as $m_\Psi\sim g_\Psi f$. The same can be applied to the KK-towers of the $SU(2)_L\cross U(1)_Y$ gauge bosons, where additional parameters $\kappa,\kappa'$ correspond to $g_\rho$, defining the mass scale of the gauge resonances as $m_\rho\sim g_\rho f$ (which further splits in case of $\kappa\neq \kappa^\prime$).}. 

As described in detail in \cite{Csaki2004}, the terms of $S_\text{IR}$ can be translated into IR BCs for the fermionic fields, i.e.
\begin{align}
	\label{eq:ModIRBCs1}
	W_\text{IR}^\dagger\evalsub{G_n(z)}{R'}&=-C_{12}W_\text{IR}^\dagger\evalat{T_{L,n}(z)}{R'}-\kappa_Rm_nR'W_\text{IR}^\dagger\evalat{T_{R,n}(z)}{R'}\\
	\label{eq:ModIRBCs2}
	W_\text{IR}^\dagger\evalsub{F_n(z)}{R'}&=+C_{12}W_\text{IR}^\dagger\evalat{T_{R,n}(z)}{R'}+\kappa_Lm_nR'W_\text{IR}^\dagger\evalat{T_{L,n}(z)}{R'},
\end{align}
where $G_n(z)=(g_n^1(z),g_n^2(z),g_n^3(z),g_n^4(z),g_n^5(z))^T$ and equivalent for the other functions. After the aforementioned Wilson line transformation in order to get rid of the Higgs vev in the bulk, these 10 field equations on the IR brane solely mix due to the nontrivial nature of $C_{12}$.

In order to implement softened symmetry breaking in the 5D theory, vector-like mass terms are introduced via couplings of the $s_L$ and $v_R$ bulk fermions with localized 4D fields
\begin{equation}
    \sRbar=\sum_n E^5_n \psibarn(x),\qquad v_L^{3,4}=\sum_nE^{3,4}_n\chin(x)
\end{equation}
as well as mixings of the $w_L$ and $w_R$ on the UV-brane. The resulting UV action reads
\begin{align}
	S_\text{UV}=\int\dd^4 x\bigg[&-i\sR\sigma^\mu\partial_\mu \sRbar- i\bar{v}_L\sigmabar^\mu\partial_\mu v_L+\frac{c_s}{\sqrt{R}}(\sR\sL+\sLbar\sRbar)\nonumber \\
	&  +\frac{c_v}{\sqrt{R}}(v_R^Tv_L+\bar{v}_L\bar{v}_R^T) + c_w(w_R^Tw_L+\bar{w}_L\bar{w}_R^T)+ h.c.\bigg]_{z=R},
	\label{eq:fullUVaction5D}
\end{align}
with $c_s$, $c_v$, $c_w$ being dimensionless new mass parameters. This action can again be translated into dynamical boundary conditions for the fermions which lead to additional constraints on the coefficients of Eq. \eqref{eq:EOMsolutionwarpedtLn}-\eqref{eq:EOMsolutionwarpedtRn} obtained by evaluation at $z=R$,
\begin{alignat}{3}
	A_n^{1,2} &= -\frac{1}{c_w}F_n^{1,2}\quad &\quad G_n^{1,2} &= \frac{1}{c_w}B_n^{1,2}\notag\\
	B_n^{3,4} &= 0 \quad &\quad G_n^{3,4} &= -\frac{m_nR}{c_v^2}F_n^{3,4}\label{eq:sMCHM5UVboundaryparameters}\\
	A_n^5 &= \frac{m_nR}{c_s^2}B_n^5 \quad & \quad F_n^5 &= 0.\notag
\end{alignat}
Implementing these relations together with the mixed equations-of-motion relations of Eq.~\eqref{eq:warpedmixedrelations} we are left with 10 coefficients $A_n^{3,4},B_n^{1,2,5},F_n^{1,2,3,4},G_n^5$, which can be fixed by the IR conditions of Eq. \eqref{eq:ModIRBCs1} and \eqref{eq:ModIRBCs2}. 

Physical solutions to the KK modes can only exist for masses $m_n$ of the KK-tower, for which this system becomes linear dependent. The last freedom is then, eventually, fixed by normalization (which is not needed explicitly in the following), which leaves us with a complete set of solutions for the KK modes \eqref{eq:EOMsolutionwarpedtLn}-\eqref{eq:EOMsolutionwarpedtRn} at every point $z$ in the bulk. The aforementioned requirement of linear dependence between the equations can be translated into the condition that the spectral function of the fermionic KK-tower, corresponding to the determinant of the $10\times10$ coefficient matrix $M_t$ covering the system of equations constructed by \eqref{eq:ModIRBCs1} and \eqref{eq:ModIRBCs2}, should vanish
\begin{equation}
    \label{eq:spectralfunctiontop}
	\rho_t(m^2_n(\xi))=\frac{\det M_t(\xi)}{\det M_t(0)} \shouldeq 0\quad\forall~n\in\mathbb{N}.
\end{equation} 
Here, $\det M_t(0)$ is the Higgs independent determinant over which is normalized in order to remove the roots resulting from fields which in our setup do not acquire a mass (e.g. the $b$ field in the $t_L$ doublet). As a phenomenological requirement we add that the mass of the zero mode $m_0$ should reproduce the SM top mass $m_t$. The mass of the lightest top partner is then defined by $m_l\equiv m_1$ and can be found numerically using Eq. \eqref{eq:spectralfunctiontop}.\par 


\subsection{The Higgs mass in holographic models}

Below, we will explore the connection between the Higgs mass and the mass of the lightest top partner in variants of the MCHM$_5$ (including also BKTs), together with the required tuning to arrive at the correct EW vev $v^2\equiv \sin^2(\tilde{v}/f)f^2=\xi f^2$. The Higgs potential in the CH framework is generated dynamically and the mass of the Higgs boson can be obtained by the second derivative of the one-loop Coleman Weinberg potential of the holographic 5D model
\begin{align}
    \label{eq:CWpotential}
	V(h)&=\sum_r\frac{N_rN_c^{(r)}}{(4\pi)^2}\int_{0}^{\infty} \dd{p}p^3\log\rho_r(-p^2) \\
	&=V_g(h)+\sum_f V_f(h)\approx V_g(h)+V_t(h)\,,
\end{align}
with $N_c^{(r)}$ the number of color-states of the contribution and $r$ summing over the different KK-towers, where $N_r=-4$ for fermionic and  $N_r=+3$ for bosonic fields.\par
The gauge contribution 
\begin{equation}
	V_g(h)=\frac{3}{(4\pi)^2}\int\dd p~p^3\qty\big[2\ln\rho_W(-p^2)+\ln\rho_Z(-p^2)]
\end{equation}
to the potential in the MCHM has already been calculated several times and can be found e.g. in \cite{Carmona2015,Csaki:2008zd}, where the spectral functions for the gauge fields yield 
\begin{equation}
	\rho_{W,Z}(-p^2)=1+f_{W,Z}(-p^2)s_h^2,
\end{equation}
with $s_h^2\equiv\sin^2(h/f)$ and coefficients
\begin{align}
	f_W(-p^2)=&\frac{ip}{2}\frac{R'}{R}\frac{1}{S(R')(C'(R')-ipR\ln(R'/R)\kappa^2S'(R'))}\\
	f_Z(-p^2)=&\frac{ip}{2}\frac{R'}{R}\Bigg[\frac{1}{S(R')(C'(R')-ipR\ln(R'/R)\kappa^2S'(R'))}\notag\\
	&+\frac{s^2_\phi}{S(R')(C'(R')-ipR\ln(R'/R)\kappa'^2S'(R'))}\Bigg]\,.
\end{align}
Moreover, similarly as for the fermions, we considered non-vanishing gauge-boson brane-kinetic terms with coefficients $\kappa,\kappa'$, given by
\begin{equation}
	S_{\rm UV}^{\rm gauge}\supset\int\mathrm{d}^4x\left[-\frac{1}{4}\kappa^2 R \ln \left(\frac{R^{\prime}}{R}\right) W^{a \mu\nu} W^a_{\mu\nu}-\frac{1}{4}\kappa^{\prime 2}R \ln \left(\frac{R^{\prime}}{R}\right) B^{\mu\nu} B_{\mu\nu}\right]_{z=R}\,,
\end{equation}
while $S(R')\equiv S_{1/2}(R')$, $C(R')\equiv C_{1/2}(R')$ as defined in Appendix \ref{sec:appwarpedsinecosine}.

The spectral function for the top-quark KK-tower $\rho_t(-p^2)$ and thus the top quark contribution $V_t(h)$ to the overall Higgs potential depends strongly on our specific setup. The spectral function for the sMCHM$_5$ is formally given by
\begin{equation}
	\rho_t(-p^2)=1+f_2(-p^2)s_h^2+f_4(-p^2)s_h^4\,,
\end{equation}
where $f_2(-p^2)$ and $f_4(-p^2)$ are very lengthy and hard to handle in an analytical manner. We will thus evaluate the fermionic potential numerically using the expression of the spectral function via the coefficient matrix given in Eq.~\eqref{eq:spectralfunctiontop} where we substitute $m_n\to ip$ (see Appendix \ref{sec:appwarpedsinecosine} for the changes in the warped trigonometric functions as well as Appendix \ref{sec:appHiggsPotential} for the numerical realization). Implementing the observational  requirements
\begin{equation}
	\label{eq:fermionicconstraints}
	m_0\equiv m_t(f)\sim\SI{150}{\GeV}\qc\eval{\pdv{V(h)}{h}}_{h=\tilde v}=0
\end{equation}
allows us to fix two of our model parameters.


\subsection{Ghost and tachyonic states for the gauge fields}
\label{sec:ghost}

To make use of the gauge and fermionic BKTs introduced before, we must ensure that no ghost states (i.e. states with negative norms) or tachyonic modes (i.e. states with negative mass squared) appear in the 4D spectrum. 

According to the analysis in Ref.\,\cite{Wojcik2018}, we shall restrict ourselves to positive BKTs for the fermion fields. We shall instead allow for negative BKTs for the gauge fields, as long as the no--ghost condition for the zero modes is satisfied. This can be derived from the normalization condition 
\begin{equation}
    N^2\int_R^{R'}\dd{z}\left(\frac{R}{z}+\kappa^{(\prime)2}R\ln\frac{R'}{R}\delta(z-R)\right)=N^2 R\ln\frac{R'}{R} (1+\kappa^{(\prime)2})\shouldeq 1,
\end{equation}
where $\kappa$ and $\kappa'$ can be treated equivalently, which requires
\begin{equation}\label{eq:noghost}
\kappa^2,\kappa'^2>-1.
\end{equation}
Negative gauge BKTs may nevertheless imply the existence of tachyonic roots for the gauge resonances. The mass scale of these problematic states is however generically much above the cutoff of the low energy effective theory for the pNGB Higgs that we are interested in\,\cite{Wojcik2018}. In the following we will then apply only the condition \eqref{eq:noghost} to our parameter space.


\subsection{Fine tuning in Composite Higgs Models}\label{subsec:tuning} 
It is well known (see e.g. \cite{Panico2013}) that generic CHMs have an inevitable tuning due to the separation of scales between EWSB and the spontaneous breaking of $SO(5)\to SO(4)$ (i.e., $v\ll f$). For $f\simeq \SI{800}{\GeV}$, which we will use throughout the paper, the parameter $\xi=v^2/f^2=\sin^2(\tilde{v}/f)\! =\!s_{\tilde v}^2 \approx 0.1$ encodes a minimal (irreducible) tuning of $\Delta=1/\xi=10$. This minimal order of tuning is, amongst others, expected for models featuring `maximal symmetry' -- which we will comment on later (see \cite{Csaki2017,Csaki2018,Blasi2020}). However, the MCHM$_5$ as well as the sMCHM$_5$ (without maximal symmetry) are expected to suffer from an enhanced tuning (`double-tuning'), due to the scalar quartic coupling being generically suppressed. Concretely, the higher trigonometric contribution to the Higgs potential $\sim \sin^4(h/f)$ appears only at subleading order in linear mixings of SM and composite fermions, which adds to the tuning to realize  $\tilde v \ll f$ by an interplay of terms \cite{Matsedonskyi:2012ym,Panico2013,Archer:2014qga,Carmona2015}.\footnote{Yet, in \cite{Blasi2020} it was shown that combining maximal symmetry and soft breaking can lead to a fruitful conflation that features very little tuning while fulfilling the stringent constraints from LHC top-partner searches.} 

In order to quantify the tuning in these models, we make use of the well established Barbieri-Giudice measure (see \cite{Barbieri1988,Csaki2017}) 
\begin{equation}
    \label{eq:tuninggeneral}
	\Delta_\text{BG}=\max_{x_i}\left|\frac{\partial\ln{\mathcal{O}}}{\partial\ln{x_i}}\right|\,,
\end{equation}
where $\mathcal{O}$ is a physical observable and $x_i$ are the free parameters of the model the observable depends on, in our case $c_\chi,c_\psi,c_1,c_2,c_s,c_v$ and $c_w$ as well as gauge and fermionic kinetic couplings. In the folowing analysis, we will calculate the fine-tuning on basis of two independent observables, the $Z$-boson mass $m_Z^2(h)$ and the mass of the physical Higgs boson $m_h^2(h)$, where
\begin{align}
	\Delta^Z_{BG}&=\max_{x_i}\left|\frac{4x_i(1-s_h^2)}{f^2m_h^2}\pdv{x_i}\left(\pdv{V(s_h^2)}{s_h^2}\right)\right|_{s_h^2=\xi}\\
	\Delta^h_\text{BG}&=\max_{x_i}\left|\frac{x_i}{m_h^2}\pdv{x_i}\left(\pdv[2]{V(h)}{h}\right)\right|_{h=\tilde v}\,,
\end{align}
and take the maximum of those for each case. 


\subsection{The mass spectrum of the sMCHM$_5$}

\begin{figure}
    \subfigure{\includegraphics[width=0.47\textwidth]{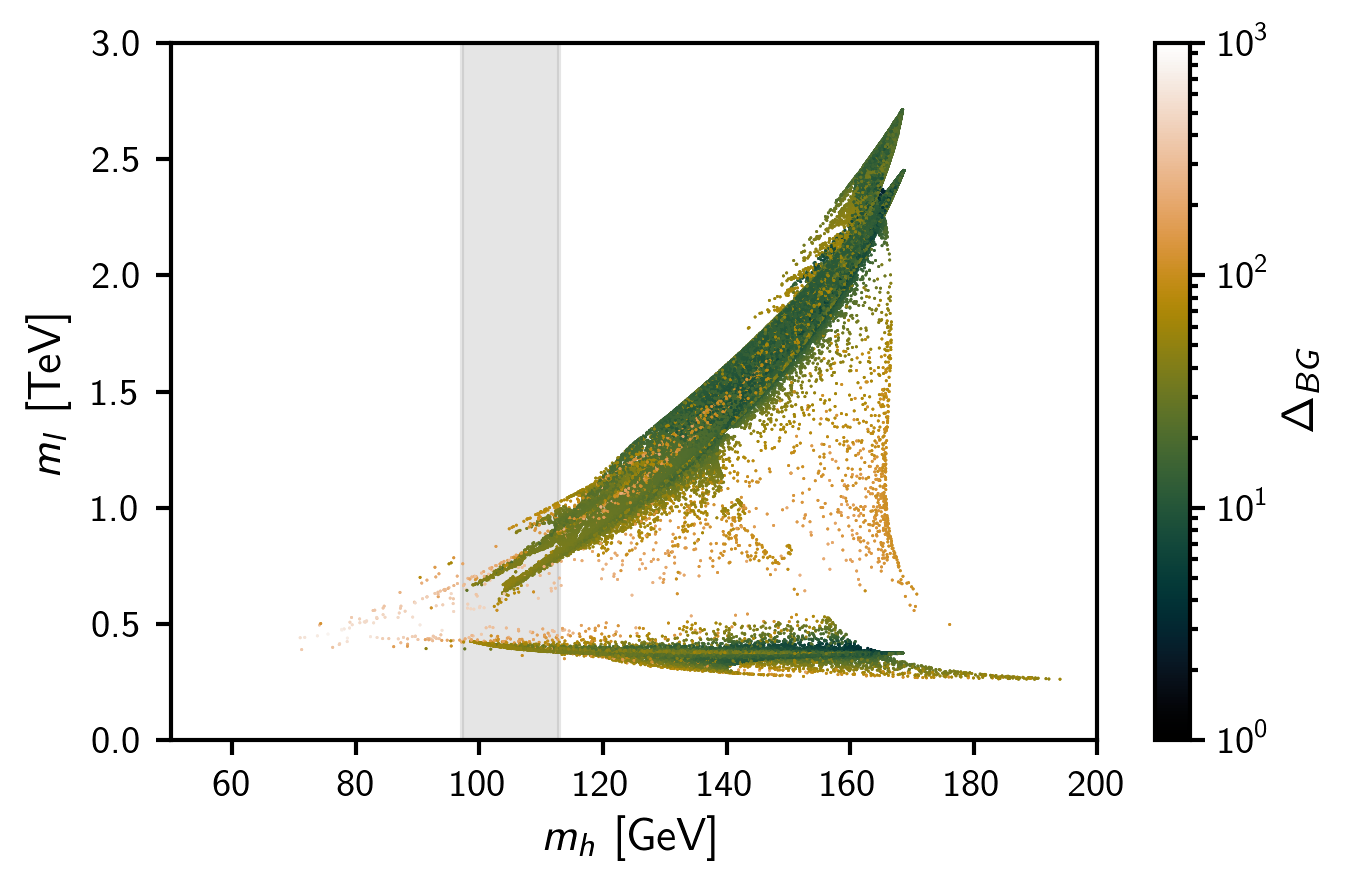}}\hspace{0.48cm}
    \subfigure{\includegraphics[width=0.47\textwidth]{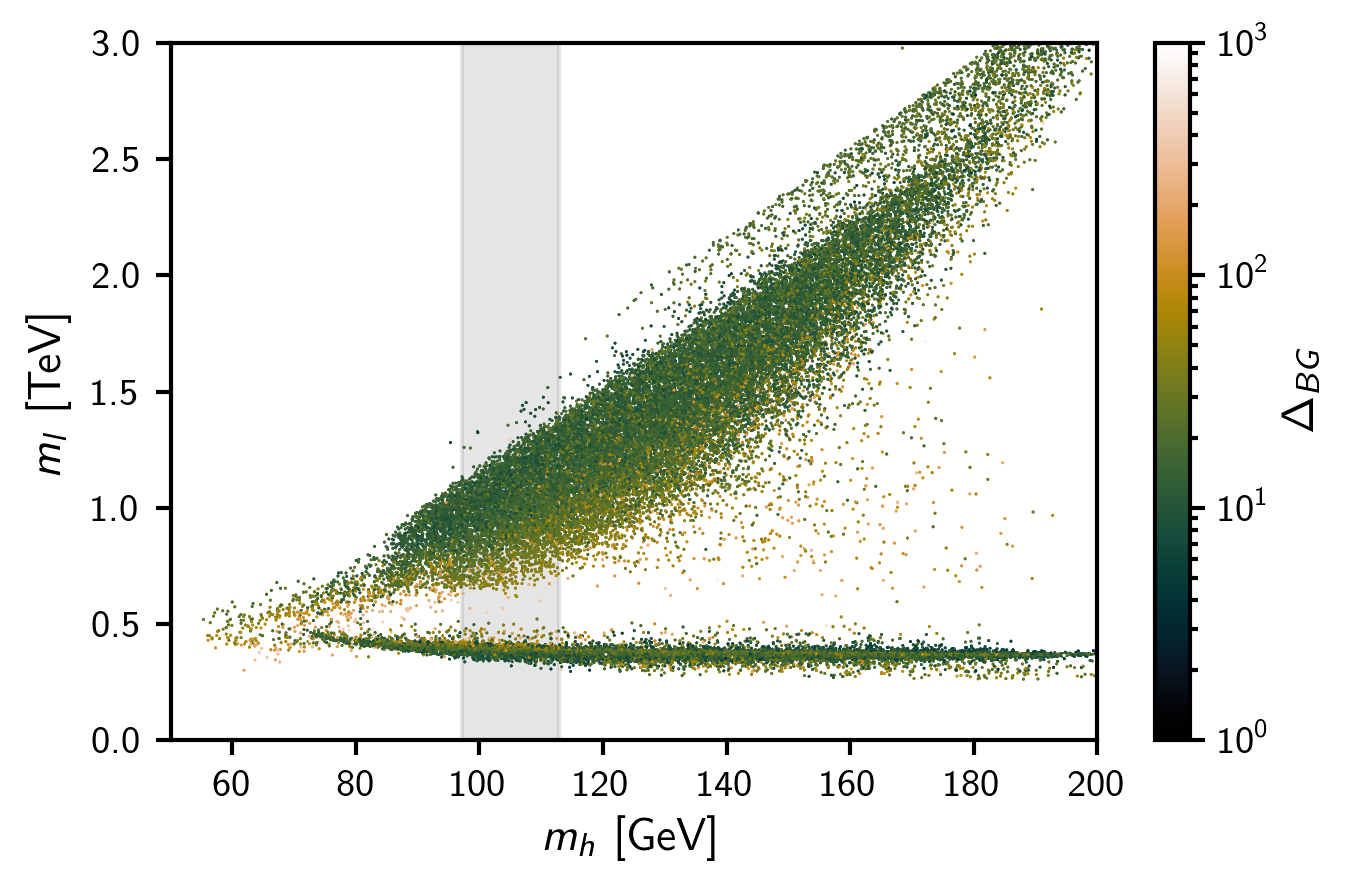}}
    \caption{The lightest top partner mass $m_l$ versus the Higgs mass $m_h$ for a large set of input parameters $c_\chi$, $c_\psi$, $c_1$ and $c_2$ in the MCHM$_5$ ($c_s,c_v,c_w \to \infty$), where $c_{\chi},c_{\psi}\in (-1/2,1/2)$ and $c_1,c_2$ fixed from Eq.~\eqref{eq:fermionicconstraints}.
    In the left plot all BKTs are set to 0 while in the right plot $\kappa^2,\kappa'^2\in [-1,1/2]$ and $\kappa_L,\kappa_R\in [0,1/2]$. The tuning for each point calculated by Eq. \eqref{eq:tuninggeneral} is color-coded, ranging from $\Delta_\text{BG}=1$ (no tuning) to $\Delta_\text{BG}=1000$ (tuning at a permille level). The gray band indicates the region of a physical Higgs mass, evaluated at the TeV scale.} \label{fig:MCHM5}
\end{figure}

We would first like to visualize the lack of sufficiently high top partner masses in standard minimal CHMs within our region of interest, i.e. for Higgs masses around $\SI{100}{\GeV}$ generated at the $\si{\TeV}$ scale. Therefore, we have illustrated in Fig.~\ref{fig:MCHM5} the limiting case of decoupled $s_R$, $v_R$ and $w$ fields (i.e. $c_s,c_v,c_w \to \infty$) which recovers the MCHM$_5$. Each point on the left plot shows the lowest resonance mass $m_l$ from a numerical solution to Eq.~\eqref{eq:spectralfunctiontop} and the Higgs mass $m_h$ from Eq.~\eqref{eq:Higgsmassnum} for a given set of input parameters $c_\chi$, $c_\psi$, $c_1$ and $c_2$ with all fermionic and gauge BKTs set to 0. Note that we have scanned $c_{\chi},c_{\psi}\in (-1/2,1/2)$ and fixed $c_{1},c_{2}$ through the observational requirements given in Eq.~\eqref{eq:fermionicconstraints}. In the right plot of Fig.~\ref{fig:MCHM5} we have allowed for non-vanishing BKTs as additional input parameters, where $\kappa^2,\kappa'^2\in [-1,1/2]$ and $\kappa_L,\kappa_R\in [0,1/2]$ according to the constraints outlined in Section \ref{sec:ghost}. 

As expected from 4D dual theories \cite{DeCurtis:2011yx,Matsedonskyi:2012ym,Pomarol:2012qf,Csaki:2008zd} and prior 5D holographic implementations (see \cite{Carmona2015}), we observe in both plots a linear proportionality between the top partner mass and the mass of the Higgs boson in the holographic MCHM$_5$, which elucidates the issue of too low top partner masses for a realistic Higgs mass (the gray bands in all figures), aimed to be resolved in this work via employing the concept of soft breaking of the Higgs shift symmetry in 5D.\footnote{The mass of the lightest top partner $m_l$ is determined by mixing of states on the IR brane. It will be highest if the dominant contribution comes from the top sector and can become even lighter if other contributions become relevant, which explains the broad spectrum as well as the upper limit of $m_l$ for a given Higgs mass.} Along the lines as discussed in Section~\ref{subsec:tuning} we can also detect an enhanced tuning $\Delta_\text{BG} \gtrsim 50$ for most input configurations, which is in agreement with our expectations. We can see from the right plot of Fig. \ref{fig:MCHM5} that while adding BKTs does give an easier access to viable Higgs masses indicated by a lower tuning within this region, it cannot significantly raise the masses of the top partners, simply because it does not break their aforementioned linear correlation with the Higgs mass.

\begin{figure}
    \subfigure{\includegraphics[width=0.48\textwidth]{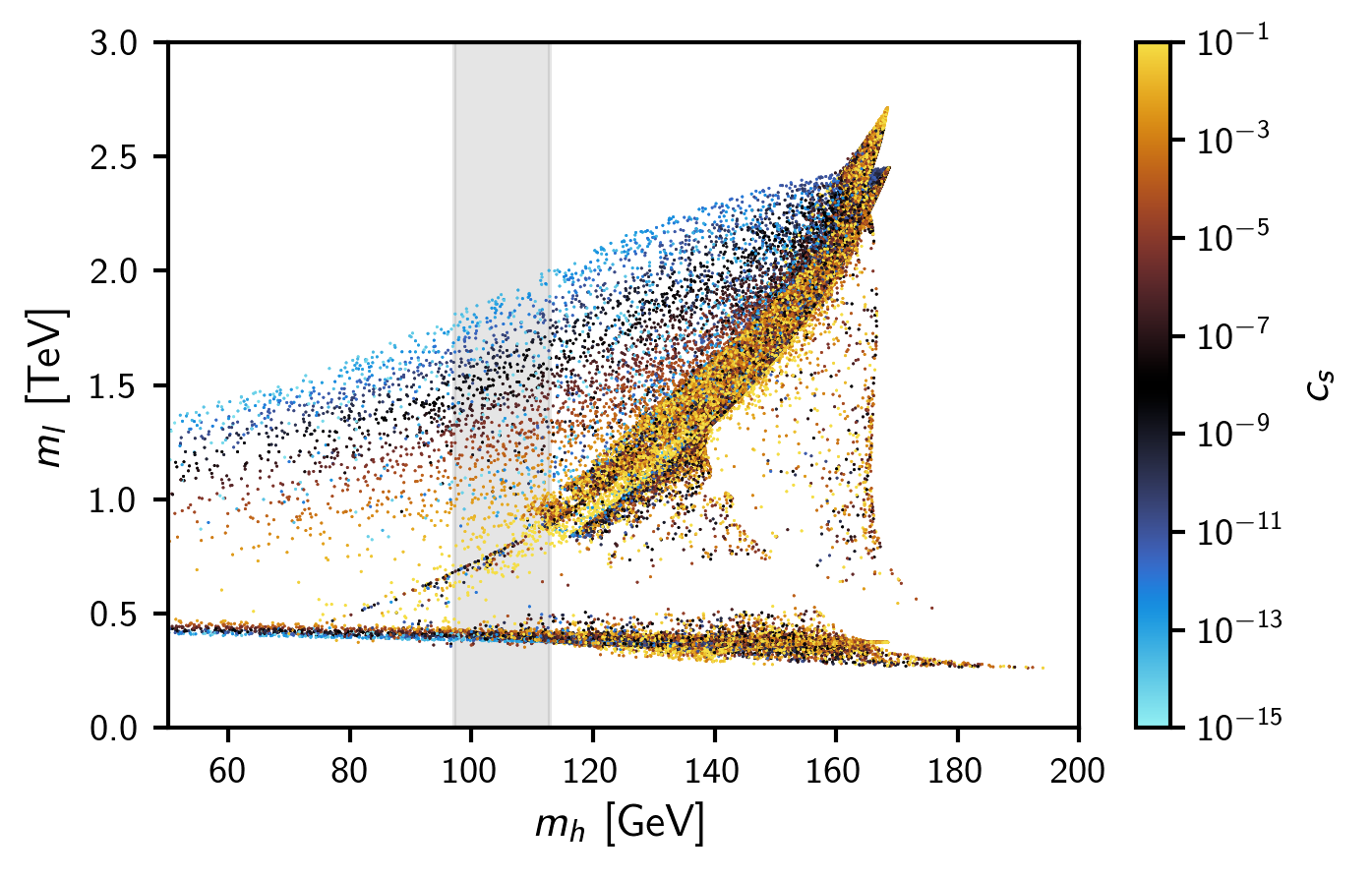}}\hspace{0.48cm}
    \subfigure{\includegraphics[width=0.47\textwidth]{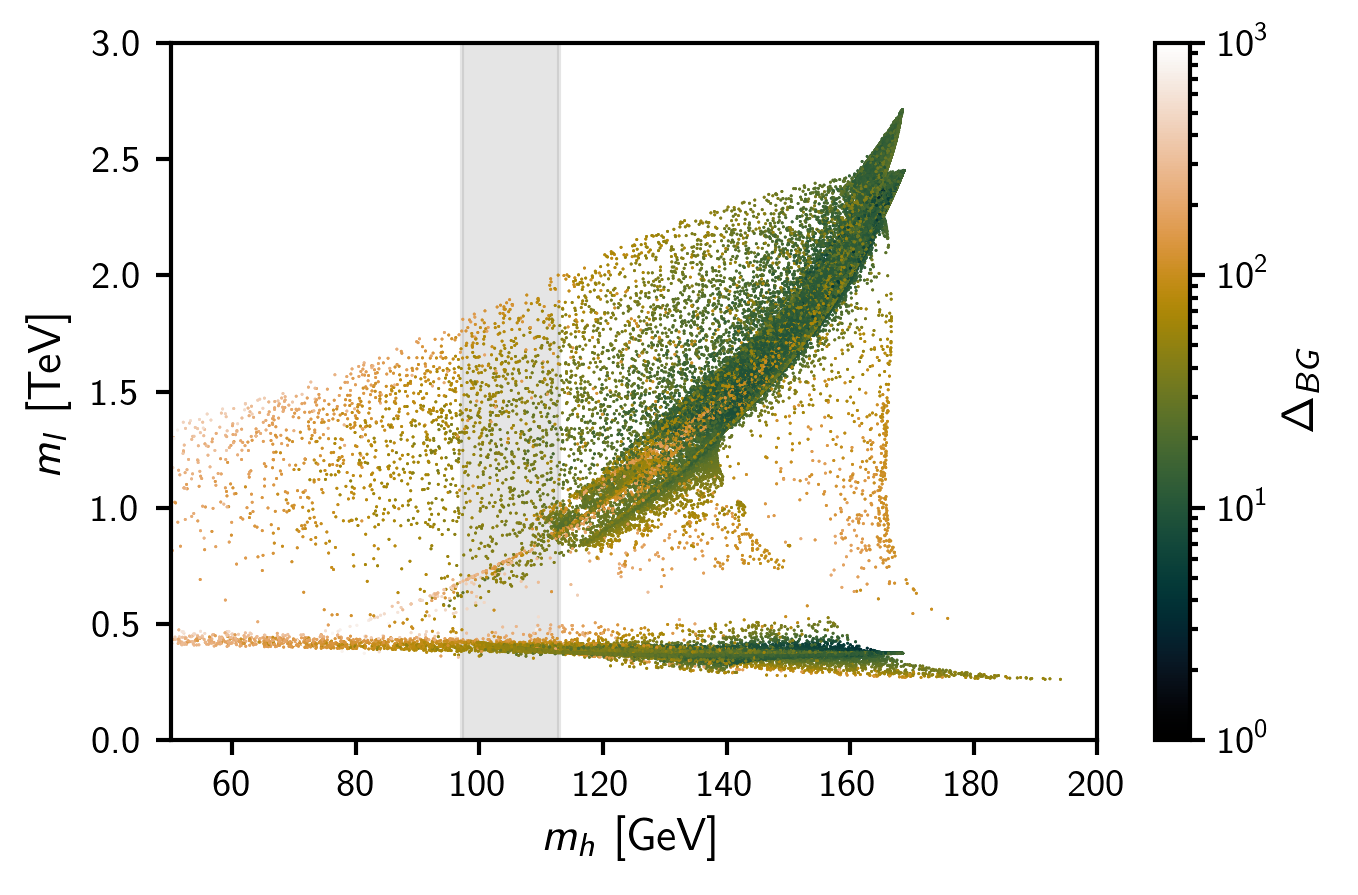}}
    \caption{The lightest top partner mass $m_l$ versus the Higgs mass $m_h$ for the MCHM$_5$ with an additional singlet $s_R$ ($c_v,c_w \to \infty$). We use a set of points as in Fig.~\ref{fig:MCHM5} with all BKTs set to $0$ and $c_s\in [10^{-15},0.1]$. In the left (right) plot, the dimensionless mass parameter $c_s$ of the vector-like singlet $s$ (the tuning) is color-coded.}
    \label{fig:sRMCHM5}
\end{figure}

To get a clear and simple view on the impact of softened symmetry breaking, we start our investigation by just including the singlet $s_R$ into the sMCHM$_5$ (i.e. $c_v,c_w\to\infty$). This simplified model, from now on called $\sR$MCHM$_5$ already removes part of the hard breaking of the global Goldstone symmetry and exhibits the desirable features of softened breaking, as depicted in Fig. \ref{fig:sRMCHM5}. We can immediately see that the strong correlation between the Higgs and the top partner mass is lifted and the Higgs mass gets reduced due to the softened breaking, entering the viable region for $m_l$ as large as 2\,TeV, i.e. twice as heavy as in the original MCHM$_5$, for a moderate tuning of $\Delta_\text{BG} \sim 100$ (indicated in the right plot). This agrees relatively well with the findings in the 4D approach, presented in \cite{Blasi2019}, even though the somewhat larger effects in the latter indicate that with more model building there might still be room for further improvement also in the holographic setup. 

We also note that the uplifting of the partner masses strongly correlates with the mass of the vector-like singlet which limits further enhancement. This is illustrated in the left panel of the figure, where the UV-boundary mass of the singlet is indicated via the color-coding. The physical mass for the singlet within an IR localized fermion field prior to mixing is given by \cite{Blasi2020}
\begin{equation}
\label{eq:singlet mass}
    m_s^2\sim \frac{(1-4c_\psi^2)c_s^2}{\abs{1+2c_\psi-c_s^2}}R'^2\left(\frac{R}{R'}\right)^{-1-2c_\psi}\quad\text{for}\quad c_\psi<1/2,
\end{equation}
such that around $c_s\sim 10^{-15}$ or equivalently $m_s\sim \order{\si{\TeV}}$, the singlet itself becomes the lightest top partner for typical values of $c_\psi$. Then, a further decrease in its mass, reducing further the global symmetry breaking,  becomes futile. 

\begin{figure}
    \subfigure{\includegraphics[width=0.48\textwidth]{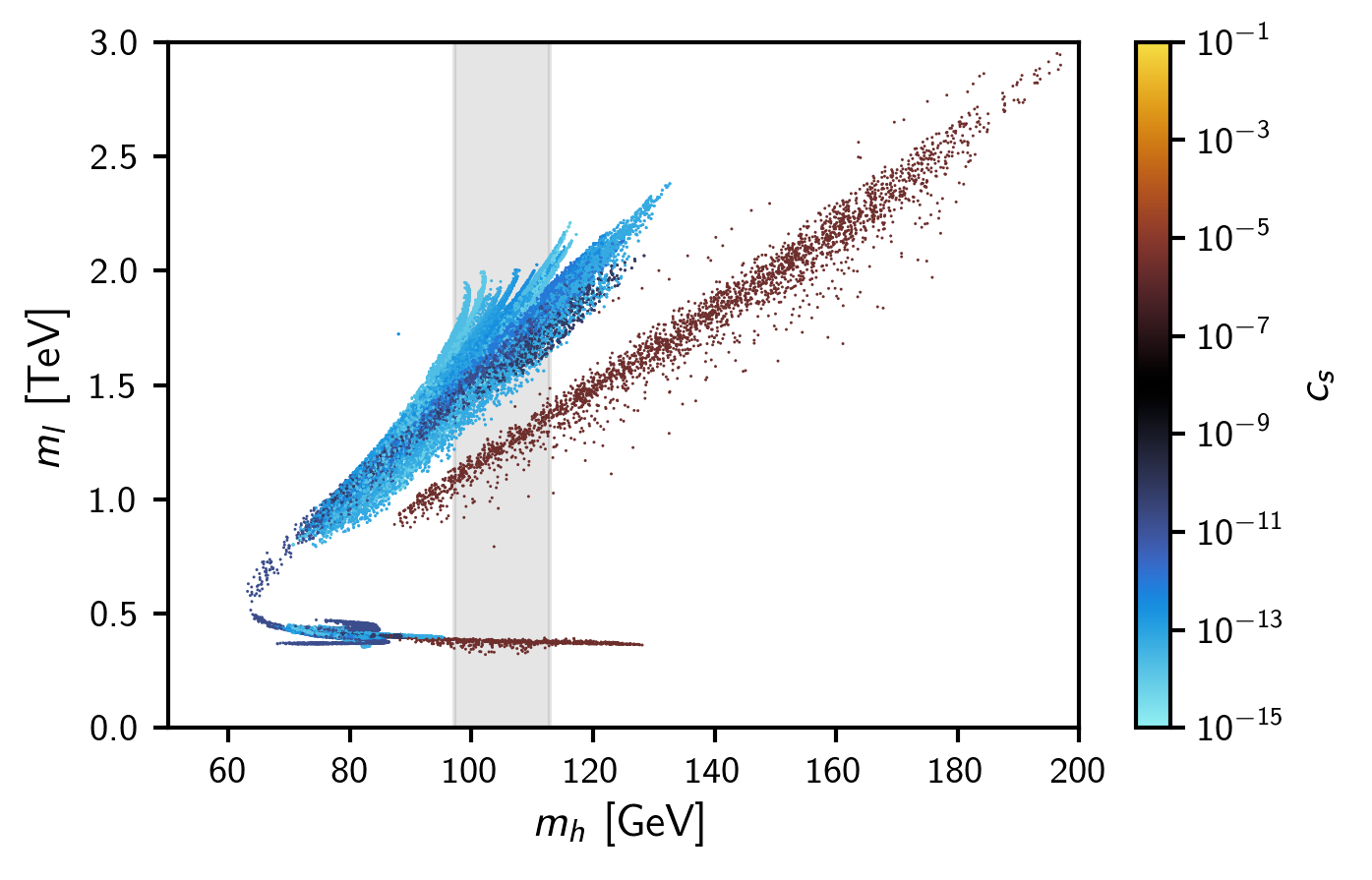}}
    \subfigure{\includegraphics[width=0.48\textwidth]{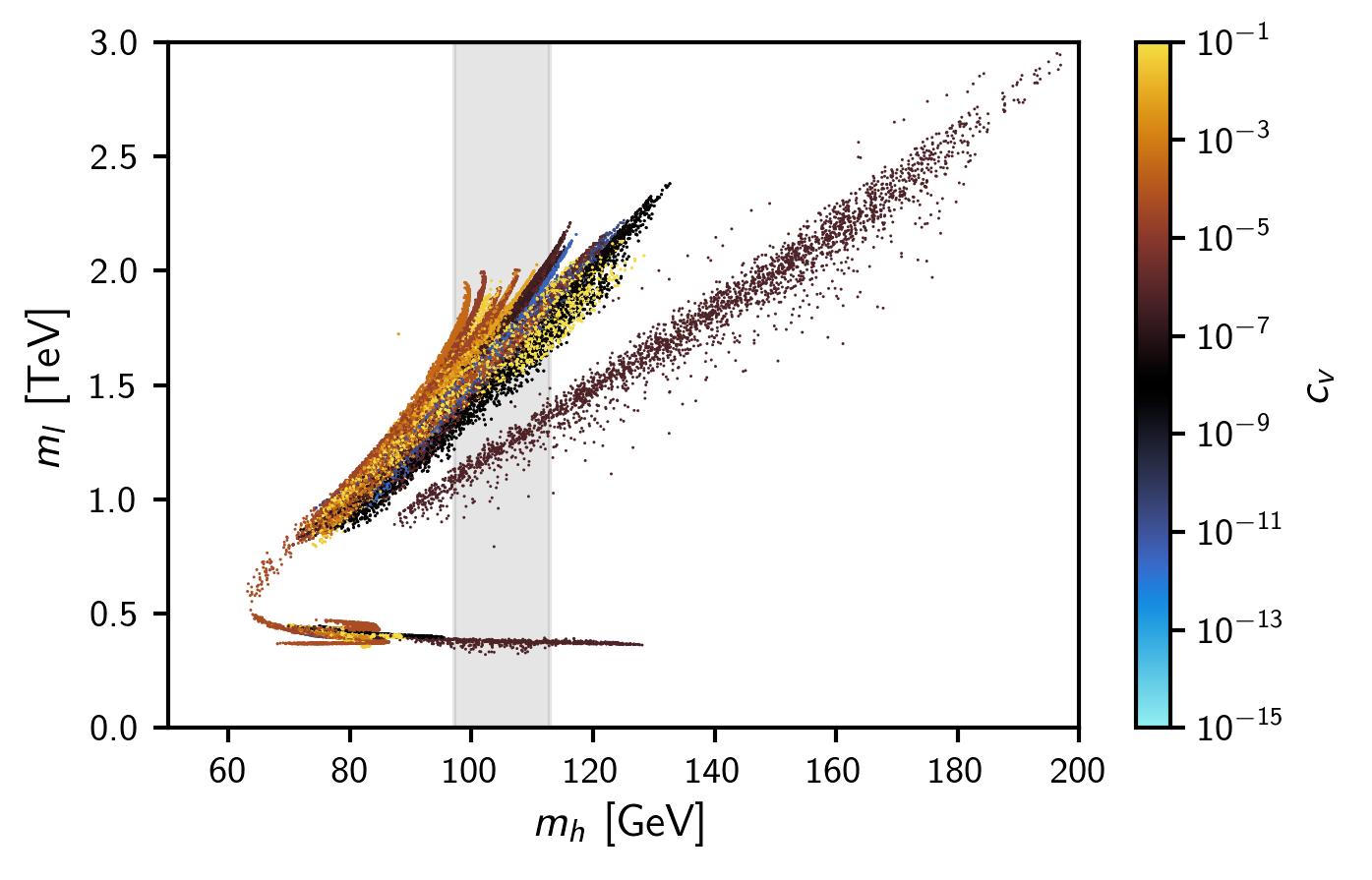}}
    \subfigure{\includegraphics[width=0.48\textwidth]{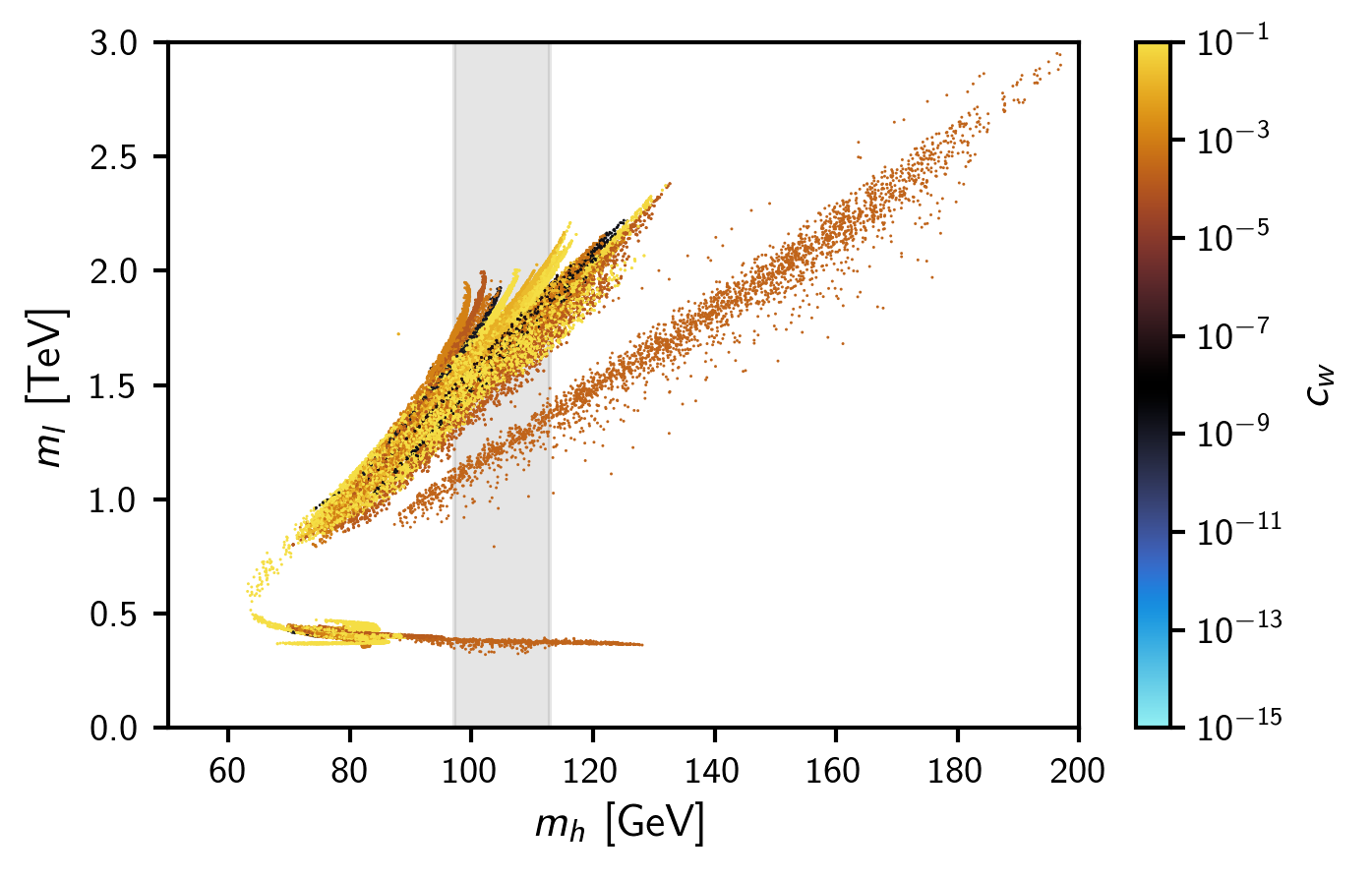}}
    \subfigure{\includegraphics[width=0.48\textwidth]{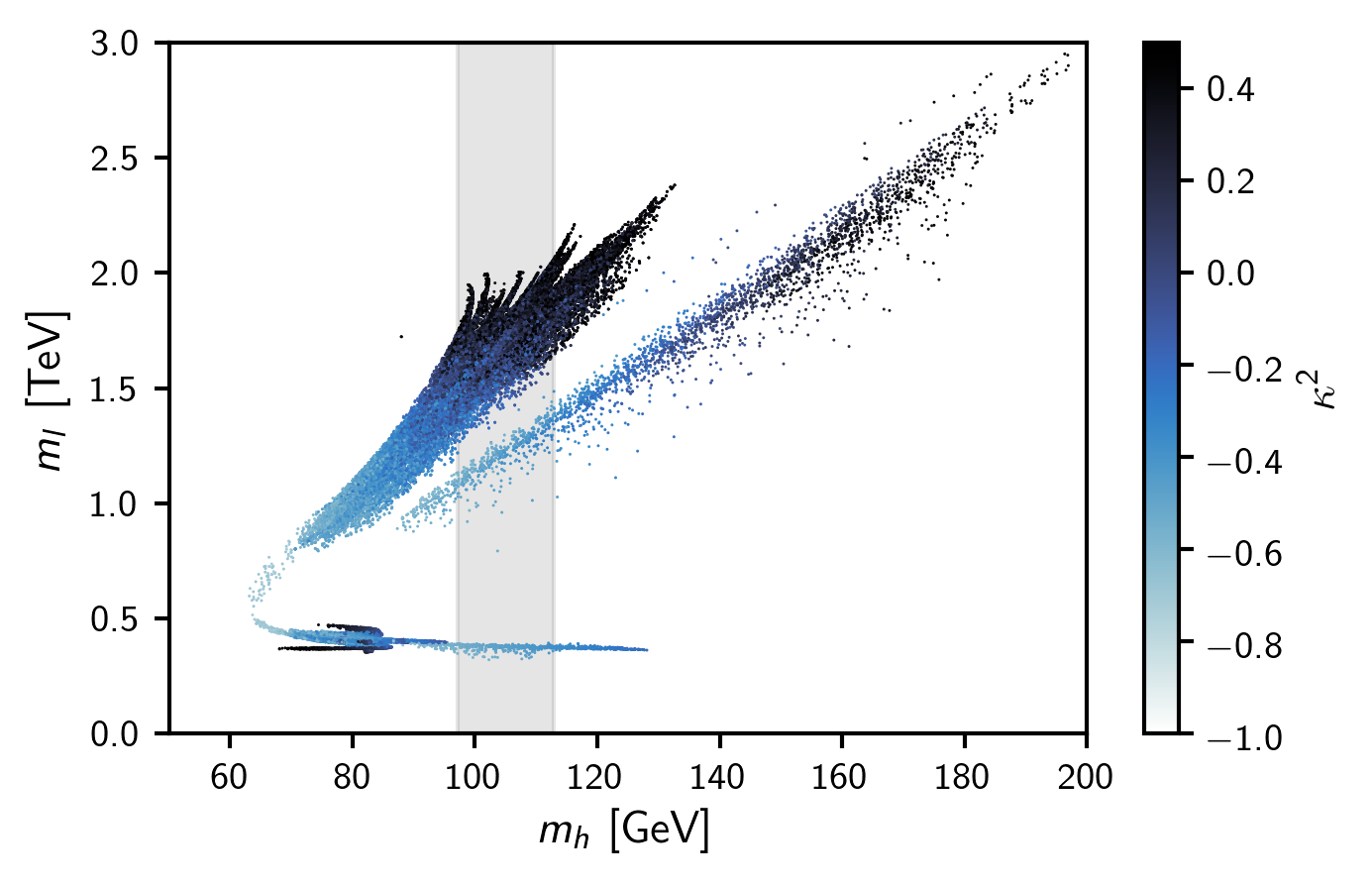}}
    \subfigure{\includegraphics[width=0.48\textwidth]{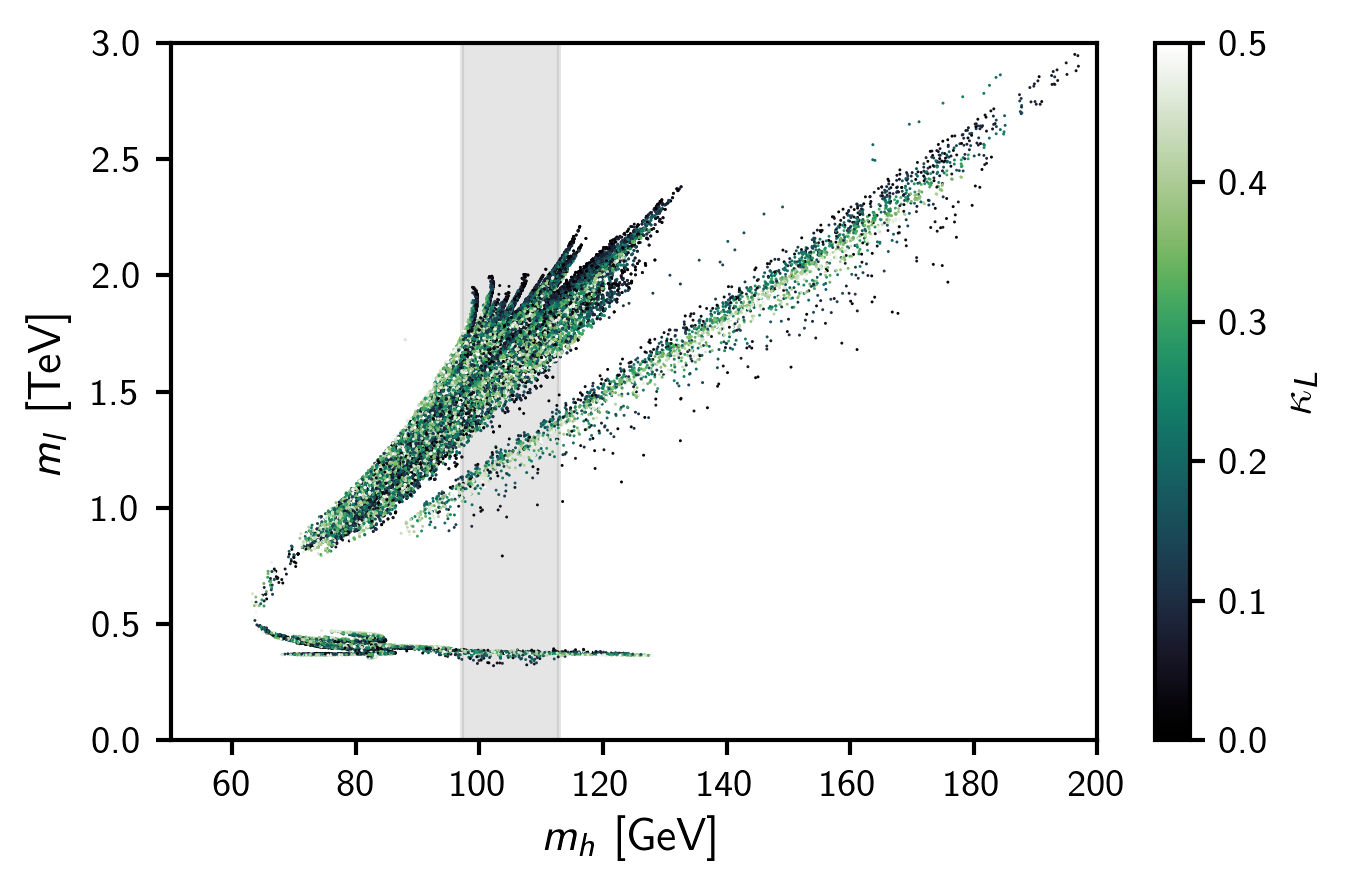}}\hspace{0.48cm}
    \subfigure{\includegraphics[width=0.47\textwidth]{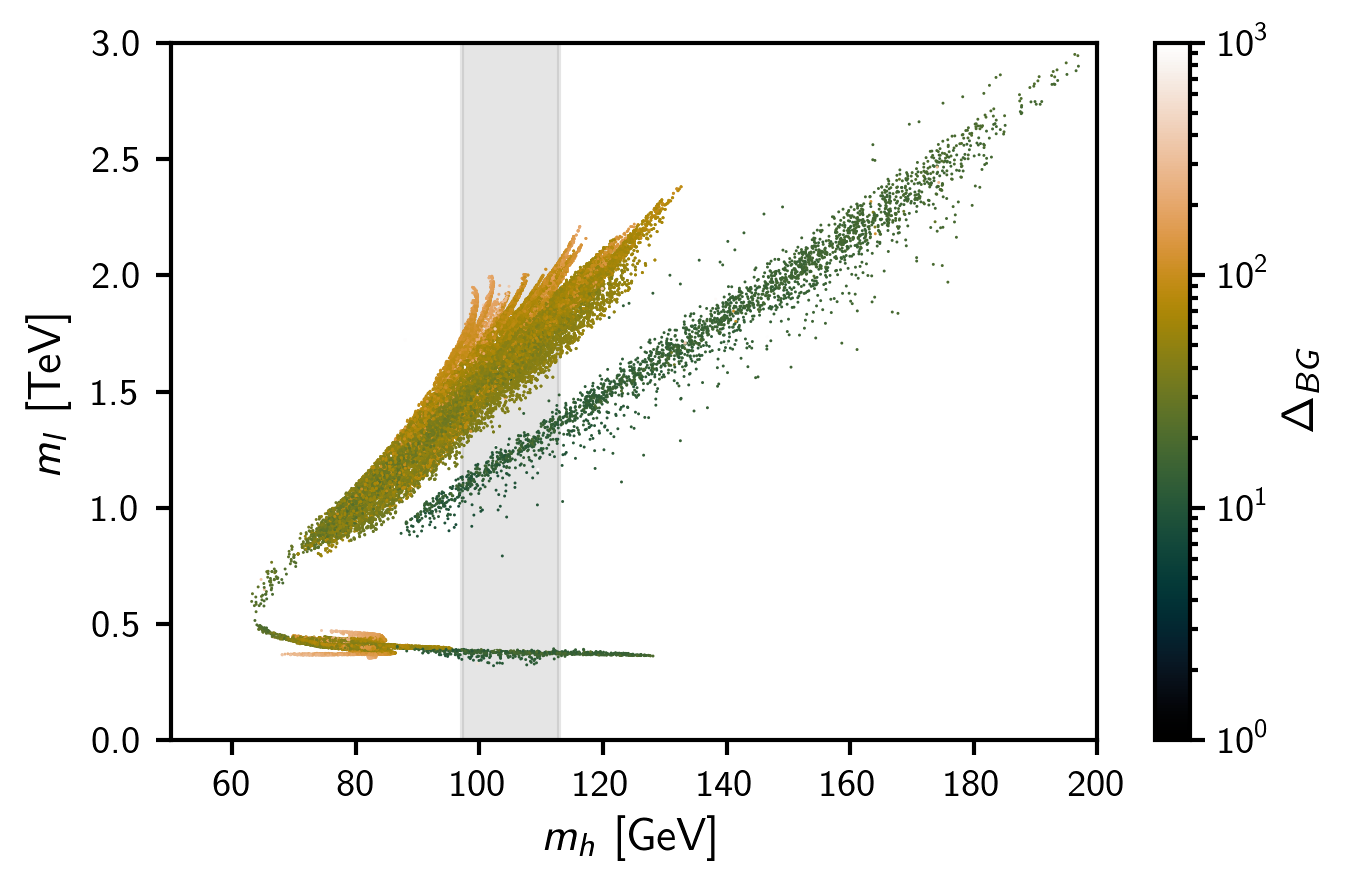}}
    \caption{The lightest top partner mass $m_l$ versus the Higgs mass $m_h$ for the full sMCHM$_5$ where the general settings are the same as in Figs.~\ref{fig:MCHM5} and \ref{fig:sRMCHM5}. We have selected $\sim 50$ random benchmark points with $c_\chi,c_\psi\in(-0.5,0.5)$, $c_1,c_2$ fixed by the conditions in Eq.~\eqref{eq:fermionicconstraints} and $c_s,c_v,c_w\in[10^{-15},0.1]$. For these, we varied $\kappa^2,\kappa'^2\in [-1,1/2]$ and $\kappa_L,\kappa_R\in [0,1/2]$. The six plots contain the same data in the $m_h-m_l$ plane, but each one color-codes a different parameter, namely $c_s,c_v,c_w,\kappa^2,\kappa_L,\Delta_{\tiny{\text{BG}}}$ from top-left to down-right.}
    \label{fig:sMCHM5}
\end{figure}

We expect this limitation to loosen in the full sMCHM$_5$ due to the interplay of the vector-like fermions. In order to verify these prospects we have performed a full analysis of the sMCHM$_5$, including fermionic and bosonic BKTs. Due to the large amount of degrees of freedom, we choose $\sim 50$ benchmark points in Fig. \ref{fig:sMCHM5}, where we only varied the fermionic and bosonic BKTs. We observe that in the sMCHM$_5$ (like in our $\sR$MCHM$_5$ toy model) we are able to access a much larger parameter space than in the MCHM$_5$. Within the region of interest, it is now possible to achieve lightest top partner masses above $\SI{2}{\TeV}$ whilst keeping the tuning moderate, of order $\Delta_\text{BG}\approx50-100$. It appears from Fig. \ref{fig:sMCHM5} that the singlet $s$ is, due to its comparably light UV-boundary mass, the driving force behind the lighter top partner states, followed by the $v$. The $w$ doublet, however, has very little impact on the SM and composite mass spectrum, as it is for most points almost decoupled from the theory.

An explanation for the obtained mass range of the new fermions can be found in their placement within the $SO(5)$ multiplets, guaranteeing the $s$ and $v$ -- which have the same quantum numbers as $t_R$ and $t_L$, respectively -- an overall bigger impact on the top sector. This is reflected by their impact on the top mass
\begin{equation}
    m_t^2\sim\frac{(c_1-c_2)^2\xi}{2R'^2\left(f_{c_\chi}^{-2}+\kappa_R+\frac{c_2^2}{1+2c_\psi}+\frac{(1-2c_\psi)c_2^2}{m_s^2R'^2}\right)\left(f_{-c_\psi}^{-2}+\kappa_L+\frac{c_1^2}{1-2c_\chi}+\frac{(1+2c_\chi)c_1^2}{m_v^2R'^2}\right)}\,,
\end{equation}
where
\begin{equation}
    f_c\equiv \sqrt{\frac{1+2c}{1-\left(\frac{R}{R'}\right)^{1+2c}}}\,,
\end{equation}
which is sizable for moderate masses $m_s$ ($m_v$), given by Eq.~\eqref{eq:singlet mass} ($c_\psi\to -c_\chi$), while the impact of the $w$ mass is of higher order in $\xi$. Regarding brane kinetics we observe that especially bosonic BKTs allow for a broader variation of the top partner masses, whereas fermionic BKTs play a subdominant role. 

After evaluation of a data set containing $\order{10^6}$ input variations, we thus confirmed that the implications of softened breaking can also mitigate the issue of light top partner masses in holographic CHMs. 

\section{Maximal symmetry from a holographic perspective}
\label{sec:maxsym}

Due to the aforementioned `double-tuning' problem in minimal CHMs it seems natural trying to alter these models in order to actually reduce the fine-tuning. One promising approach which can easily be included into the model at hand is the concept of maximal symmetry \cite{Csaki2017,Csaki2018}. The idea is to start with an enhanced chiral symmetry for the fermions, $SO(5)_L\times SO(5)_R$. For a certain relation between the masses of the composite resonances residing in the corresponding multiplets, see below, this higher symmetry is then broken into a remaining, so-called maximal symmetry $SO(5)' \supset SO(4)$ of the composite sector -- which is the maximal symmetry that does not contain the Goldstone shift symmetry any more and thus creates a potential for the Higgs. However, due to maximal symmetry the leading contribution to the quadratic scalar term vanishes, which removes the enhanced `double-tuning' of the MCHM$_5$. In practice, this symmetry emerges when the masses of the composite fourplets of $SO(4)$ are equal in magnitude but opposite in sign to those of the composite singlets, see \cite{Csaki2017,Csaki2018,Blasi2020} for more details. 

\subsection{Maximal symmetry in the holographic MCHM$_5$}

In contrast to findings from earlier work in 4D dual theories (e.g. \cite{Csaki2018}), maximal symmetry in the holographic MCHM$_5$, which we analyze here for the first time, strongly constrains the fermionic spectrum, making it more challenging to construct realistic 5D models. As we shall see, maximal symmetry implies that the Higgs mass after EWSB only depends on the parameters of the gauge sector. The starting point for our calculation is the potential from Eq. \eqref{eq:CWpotential}, which can be generally written as 
\begin{equation}
    V(h)=(\gamma_g-\gamma_f)s_h^2+(\beta_g+\beta_f)s_h^4
\end{equation}
where subscripts $g$ and $f$ indicate the gauge and fermion contributions, respectively. 

In the MCHM$_5$, maximal symmetry
implies an unbroken trigonemetric parity
in the fermionic sector,
$s_h^2\leftrightarrow -c_h^2$, which forces $\gamma_f=\beta_f$ \cite{Csaki2017,Blasi2020}. The Higgs mass after EWSB reads 
\begin{equation}
    m_h^2=\frac{8(\beta_f+\beta_g)}{f^2}(1-\xi)\xi=\frac{8(\gamma_f+\beta_g)}{f^2}(1-\xi)\xi
\end{equation}
where $\xi=s_{\tilde v}^2$, i.e. evaluated at the minimum $h=\tilde v$. On the other hand, the minimization condition becomes
\begin{equation}
\label{eq:minC}
    0=\pdv{V}{s_h^2}=\gamma_g-\gamma_f+2(\gamma_f+\beta_g)\xi\quad\Rightarrow\quad \gamma_g=(1-2\xi)\gamma_f-2\beta_g\xi\,,
\end{equation}
exhibiting that the gauge contribution is crucial to allow for a viable $\xi \neq 1/2$ \cite{Csaki2017,Csaki2018}.
Using these two conditions we arrive at 
\begin{equation}
\label{eq:gammagmax}
    \gamma_g=\frac{1-2\xi}{8(1-\xi)\xi}f^2m_h^2-\beta_g
\end{equation}
or equivalently
\begin{equation}
    m_h^2=\frac{8(1-\xi)\xi}{1-2\xi}\left(\frac{\gamma_g+\beta_g}{f^2}\right)\simeq \frac{8(1-\xi)\xi\gamma_g}{(1-2\xi)f^2}\,,
\end{equation}
where we have used $\gamma_g\gg\beta_g$. 

In 4D language, $\gamma_g$ is given by \cite{Marzocca:2012zn} 
\begin{equation}
    \gamma_g\approx \frac{9}{64\pi^2}g^2f^4g_\rho^2
\end{equation}
with $g$ the usual EW gauge coupling and $g_\rho$ the coupling for gauge resonances (see below). Setting $\xi=0.1$, $m_h=\SI{0.12}{\TeV}$ and $v=\SI{0.246}{\TeV}$, the condition above requires a rather small $g_\rho\approx 3.27$, which is a completely fine value in the 4D theory. We note that this still induces relatively light top partners, due to the fermionic contribution still entering via the relation~\eqref{eq:minC}, requiring the latter to be also rather small which leads to light partners as in the regular MCHM$_5$.

\begin{figure}
    \subfigure{\includegraphics[width=0.48\textwidth]{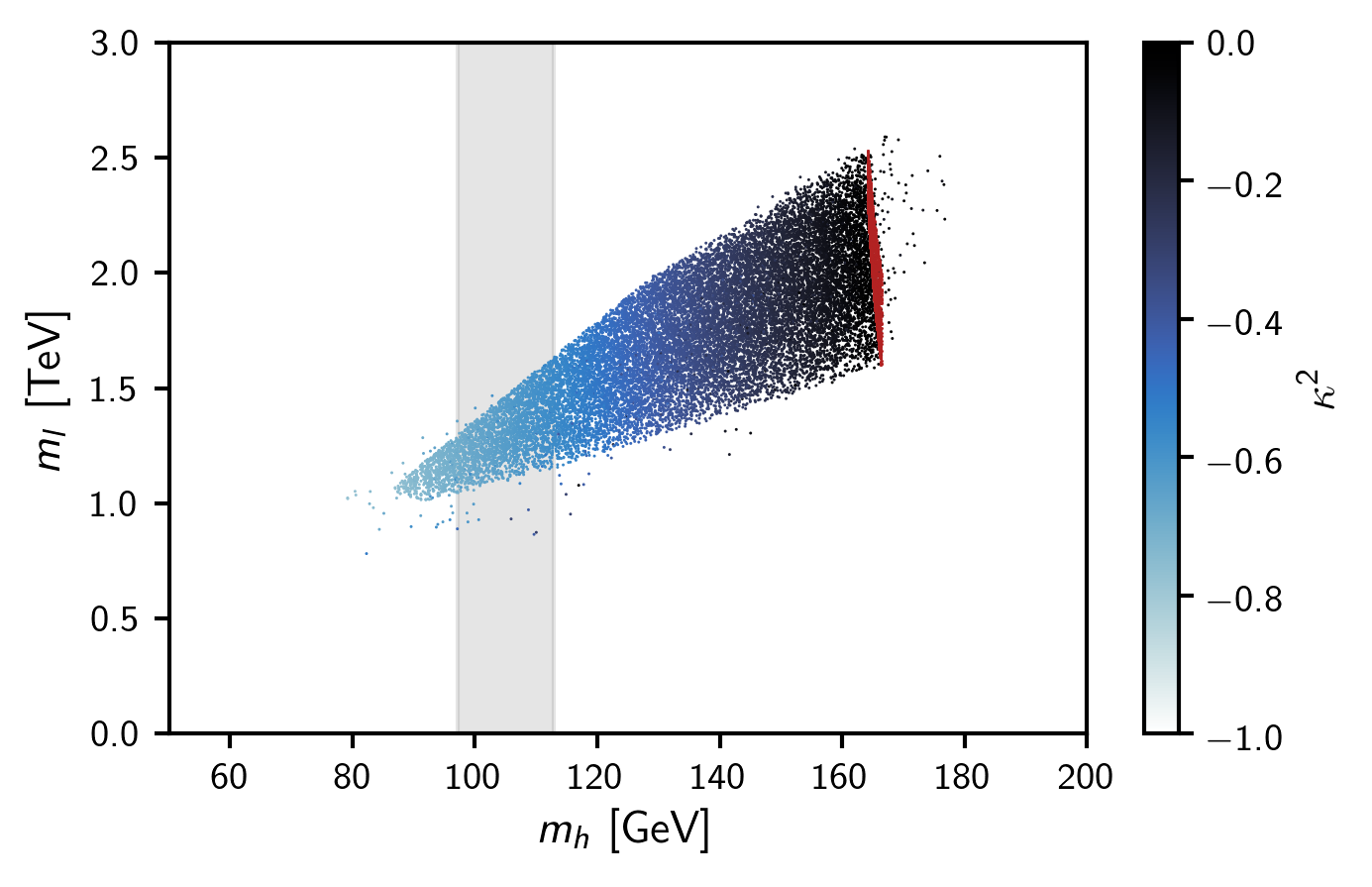}}
    \subfigure{\includegraphics[width=0.48\textwidth]{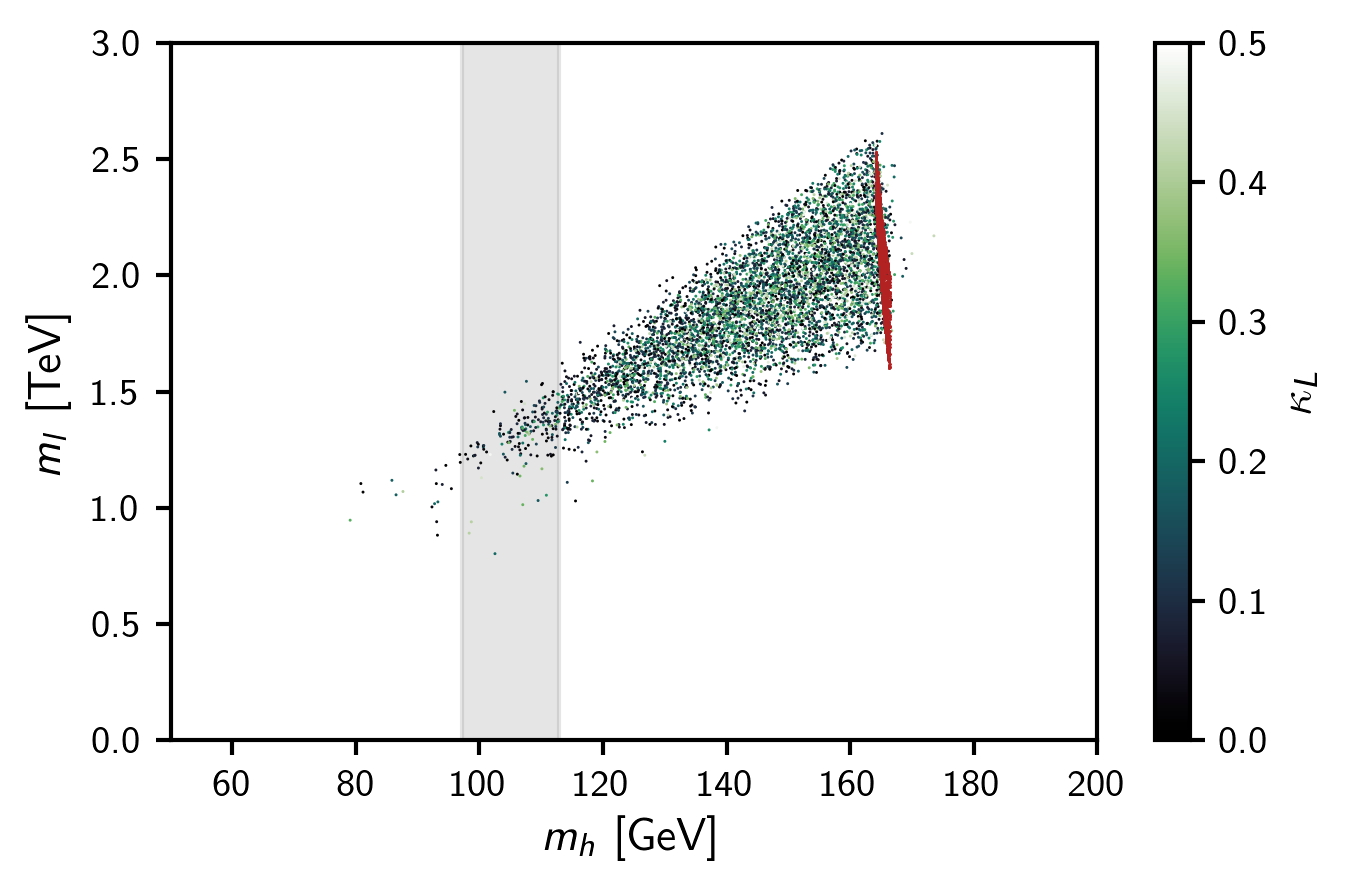}}
    \caption{The lightest top partner mass $m_l$ versus the Higgs mass $m_h$ for the MCHM$_5$ with the additional constraint $c_1=-c_2$, which imposes maximal symmetry. Whilst $c_\chi,c_\psi\in(-0.5,0.5)$, the parameter $c_1$ is fixed by the minimum condition of the Higgs potential at its physical vev. In order not to over-constrain the system, we allowed for a range of top masses between $m_t\sim 150-170\,\si{\GeV}$. In the left plot, we have allowed only for gauge BKTs $\kappa^2,\kappa'^2\in[-1,0]$ at $\kappa_L,\kappa_R=0$, whereas in the right plot we additionally also varied $\kappa_L,\kappa_R\in[0,1/2]$. Solutions with all BKTs put to $0$ are shown in red in both plots.}
    \label{fig:MCHM5MaxSym}
\end{figure}

Things are different for maximal symmetry in a minimal 5D setup, which we can easily implement for the MCHM$_5$ by setting $c_1=-c_2$ in the IR mass terms of \eqref{eq:SIR}, mirroring the 4D trigonometric parity \cite{Csaki2018,Blasi2020}. Since the 5D coupling constant \cite{Carmona2015}
\begin{equation}
    g_5^2\approx \frac{e^2}{\sin^2(\theta_W)}R\ln\left(\frac{R'}{R}\right) (1+\kappa^2)
\end{equation}
is tied to the IR radius $R'$ as well as gauge brane kinetics,
without adding BKTs ($\kappa^2=0$) and setting $f=\SI{800}{\GeV}$ (or equivalently $\xi=0.1)$, $R'=\SI{0.625}{\TeV^{-1}}$ is required from Eq. \eqref{eq:fRprimeconnection}. This fixes the value of~\cite{Carmona2015}
\begin{equation}
    g_\rho\approx 1.2024 g_*\approx4.8\,,
\end{equation}
leading to approximately $m_h\simeq \SI{0.179}{\TeV}$.  The actual calculation for $g_\rho$ in the 5D setup yields $m_h\simeq\SI{0.163}{\TeV}$ which corresponds to $m_h\simeq\SI{0.197}{\TeV}$ once evaluated at the 100\,GeV scale. This strict value for the Higgs mass following from maximal symmetry poses a problem on the validity of the minimal 5D model as is and requires modifications.  We have confirmed this issue by running simulations in our 5D setup, observing the Higgs mass to be constrained to the stated value as can be seen from the red dots in Fig.~\ref{fig:MCHM5MaxSym}

The problem can be resolved by introducing quite sizable negative gauge BKTs into the 5D model, discussed before, which shift $g_*$ to smaller values. As in the 4D theory, this however implies lower top partner masses due to a smaller $\gamma_f$ fixed by the minimization condition \eqref{eq:minC} -- see the left-hand side of Fig. \ref{fig:MCHM5MaxSym} for the numerical evaluation (exhibiting in particular the required ${\cal O}(1)$ gauge BKTs). The issue of light partners can also not be overcome by fermionic BKTs, see right panel of Fig.~\ref{fig:MCHM5MaxSym}, so to lift the top partners $f$ needs to be raised, increasing again the tuning of the setup. 
This concludes our study of the explicit holographic realization of a maximally symmetric MCHM$_5$, finding that viable models require non-negligible gauge BKTs.


\subsection{The maximally symmetric sMCHM$_5$}

In order to potentially address the issue of light top partners in the maximally symmetric MCHM$_5$ and allow for more freedom in the model building, we can straightforwardly embed maximal symmetry into a softly broken version of the setup, the sMCHM$_5$, by setting $c_1=-c_2$ in this model on the IR-brane and changing our couplings on the UV-brane. The latter is necessary because whenever we couple $\psi_L$ and $\psi_R$ directly, we fully break the trigonometric symmetry of the model and reintroduce double-tuning (see \cite{Blasi2020} for the details). The simplest possibility is to introduce two additional UV-localized 4D Weyl-spinors $\tilde{w}_R$ and $\tilde{w}_L$ to couple to the $w_L$ and $w_R$ on the UV-brane. The UV action changes into
\begin{align}
	S_\text{UV}=\int\dd^4 x\bigg[&-i\sR\sigma^\mu\partial_\mu\sRbar-i\bar{v}_L\sigmabar^\mu\partial_\mu v_L-i\tilde{w}_R^T\sigma^\mu\partial_\mu \bar{\tilde{w}}_R^T-i\bar{\tilde{w}}_L\sigmabar^\mu\partial_\mu \tilde{w}_L\nonumber \\
	&+\frac{c_s}{\sqrt{R}}(\sR\sL+\sLbar\sRbar)+\frac{c_v}{\sqrt{R}}(v_R^Tv_L+\bar{v}_L\bar{v}_R^T)\nonumber \\
	&+ \frac{c_{w_R}}{\sqrt{R}}(w_R^T\tilde{w}_L+\bar{\tilde{w}}_L\bar{w}_R^T)+\frac{c_{w_L}}{\sqrt{R}}(\tilde{w}_R^Tw_L+\bar{w}_L\bar{\tilde{w}}_R^T)\bigg]_{z=R},
	\label{eq:UVactionMaxSym}
\end{align}
with $c_{w_L}$, $c_{w_R}$ as new mass parameters of the model. This of course also changes the BCs on the UV-brane for the first two pairs into
\begin{equation}
	A_n^{1,2} = \frac{m_nR}{c_{w_L}^2}B_n^{1,2}\qquad G_n^{1,2} = -\frac{m_nR}{c_{w_R}^2}F_n^{1,2}.
\end{equation}
Additionally, a $Z_2$ symmetry is introduced, under which the $\Psi_1$ and the right-handed UV-localized fields have negative and the $\Psi_2$ along with the left-handed UV-localized fields positive parity. This forbids unwanted trigonometric-parity breaking terms if it is imposed everywhere, except on the IR-brane, where the symmetry is broken in order to give mass to the SM fermions (Eq.~\eqref{eq:SIR}).

Soft breaking will also break trigonometric parity in a controlled way -- without reintroducing the double-tuning -- due to the new vector-like masses \cite{Blasi2020}. From a 4D point of view, it can be derived that the breaking of trigonometric parity is given by
\begin{equation}
    \beta_f-\gamma_f=-C^2(m_v^2-m_{w_R}^2)
\end{equation}
with $C^2$ a positive prefactor, which modifies Eq.~\eqref{eq:gammagmax} into
\begin{equation}
\label{eq:mH_trigb}
    \gamma_g\simeq \frac{1-2\xi}{8(1-\xi)\xi}f^2m_h^2+C^2(m_v^2-m_{w_R}^2)\,,
\end{equation}
where we have again neglected $\beta_g \ll \gamma_g$. We observe that employing soft breaking, first, the new contributions allow for more freedom such that a viable $\xi \neq 1/2$ becomes easier to achieve.
Moreover, as we can see from the relation above, the Higgs can now be lighter with the same $\gamma_g$, if the second term in the RHS
is large and positive, which suggests to consider $m_{w_R}/m_v\ll1$. Actually, it is possible to show that when $m_{w_R}\to 0$, one
approaches a configuration with $-\gamma_f+\beta_f=-\beta_f$, which would result in $m_h\sim \SI{0.12}{\TeV}$ also for $\xi=0.1$ and $g_\rho\approx 4.8$, corresponding to $\kappa=0$ in the 5D model. Even though this strict limit is not of much use as a new light quark is unavoidably required in the spectrum, the above discussion shows that in the maximally symmetric sMCHM$_5$ the tight relations of the original proposal of maximal symmetry can be lifted.

One could thus hope that for finite values of the vector-like masses some cancellation of fermionic and gauge contributions in \eqref{eq:mH_trigb} could make the Higgs mass reach its physical value without necessarily lowering $\gamma_f$. Numerically, it turns out that this effect is however rather limited and still both $\gamma_g$ and $\gamma_f$ need to be reduced, as before. Still, due to softened breaking, a taming of the fermionic contributions (smaller $\gamma_f$) becomes now possible  while keeping the partner masses {\it constant}, thereby avoiding light top partners (without increasing the tuning) -- as was already found in \cite{Blasi2020} considering a 4D setup.

On the other hand, achieving the same lowering of $\gamma_f$ in 5D incarnations turns out to be not so trivial. In the straightforward 5D analogue of the setup studied here, varying the input parameters (including gauge and fermionic BKTs) within their allowed boundaries, a significant reduction of $\gamma_f$ turns out to be possible again only {\it together} with the presence of light top partners.
Indeed, we have scanned the parameter space in the proposed model and discovered no significant enhancement of top-partner masses compared to the maximally symmetric MCHM$_5$. 
A probable explanation is that the simple holographic implementation considered here provides no exact realization of the strong dynamics found to lift the top partners in \cite{Blasi2020}, where the breaking of the Goldstone symmetry in the fermionic sector could be minimized via an appropriate fermionic spectrum with also the new elementary states being sufficiently heavy. Entering the corresponding parameter space was apparently not possible in our numerical scans due to the restrictions of the particular 5D model, containing only a few fundamental parameters determining the properties of the fermions. 

Deeper modifications of the 5D setup would be required to reproduce the spectrum of composite fermionic states and elementary fermions that allowed to significantly reduce the Goldstone symmetry breaking in \cite{Blasi2020} such as to arrive at a light composite Higgs at minimal tuning without anomalously light top partners -- which invites for further studies.


\section{Conclusions}
\label{sec:conc}

We presented viable implementations of soft-symmetry breaking and maximal symmetry in 5D holographic CHMs.
For the former we showed how universal BCs for the fermion fields, combined with breaking of the global symmetry via soft brane-localized mass terms, allows to lift the strong correlation between the Higgs mass and the top-partner masses. We thus achieved to realize a light Higgs, while keeping the top partners heavier than in conventional models. This improves the prospects to still probe a non-trivial part of the spectrum of CHMs at the LHC or early FCC.

For the latter, we pointed out that maximal symmetry, while removing the double-tuning in the Higgs potential, in its minimal realization makes a sharp prediction for the Higgs mass at 197\,GeV. The strong relation between $m_h$ and the top partner masses remains present, while it is possible to arrive at a viable Higgs mass, with sizable negative gauge BKTs. 

In order to combine maximal symmetry and softened breaking of the Higgs shift symmetry in a way that leads to a minimal tuning without light top partners, it would be interesting to extend the 5D warped setup considered here. Like this, it should be possible to realize the strong dynamics and resulting fermion spectrum that minimizes the breaking of the Goldstone symmetry, as found in \cite{Blasi2020}.


\section*{Acknowledgments}

This paper is partly based on the Master's thesis of Julian Bollig. SB is supported by FWO-Vlaanderen through grant number 12B2323N. JB is suported by the Research Training Group RTG2044 funded by the German Research Foundation (DFG).


\section*{Appendix}
\appendix


\section{Dirac matrices in 5D and supersymmetric notation}
\label{sec:appDiracMatrices}
\noindent
The 5D gamma matrices on an AdS$_5$ with signature $(+,-,-,-,-)$ are given by
\begin{equation}
	\Gamma^M=\{\gamma^\mu,i\gamma^5\}\,,
\end{equation}
with $\mu=0,1,2,3$, and $\gamma^\mu$, $\gamma^5$ being the 4D gamma matrices in Weyl representation
\begin{equation}
	\gamma^\mu=
	\begin{pmatrix}
		0 & \sigma^\mu\\
		\sigmabar^\mu & 0
	\end{pmatrix}
	\qc \gamma^5=
	\begin{pmatrix}
		\mathbbm{1}_2 & 0 \\
		0 & -\mathbbm{1}_2
	\end{pmatrix},
\end{equation}
where $\sigmabar^\mu=(\sigma^0,-\sigma^k)$. The $\sigma^\mu$ are the usual Pauli matrices 
\begin{equation}
	\sigma^0=-\mathbbm{1}_2\qc \sigma^1=
	\begin{pmatrix}
		0 & 1 \\
		1 & 0 \\
	\end{pmatrix}
	\qc \sigma^2=
	\begin{pmatrix}
		0 & -i \\
		i & 0
	\end{pmatrix}
	\qc \sigma^3=
	\begin{pmatrix}
		1 & 0 \\
		0 & -1 
	\end{pmatrix}.
\end{equation}
In this basis a 5D Dirac spinor can be written as
\begin{equation}
	\Psi=\colvec{2}{\Psi_L}{\Psi_R}=\colvec{2}{\chi_\alpha}{\psibar^{\dot{\alpha}}},
\end{equation}
with $\Psi_{L,R}=\frac{1}{2}(\mathbbm{1}_4\pm\gamma^5)\Psi$, where $\alpha$ and $\dot{\alpha}$ denote supersymmetric indices. Using the $2\cross 2$ anti-symmetric tensors $\epsilon_{\alpha\beta}=(i\sigma^2)_{\alpha\beta}$, $\epsilon_{\dot{\alpha}\dot{\beta}}=(i\sigma^2)_{\dot{\alpha}\dot{\beta}}$ and their inverses $\epsilon^{\alpha\beta}=(-i\sigma^2)_{\alpha\beta}$, $\epsilon^{\dot{\alpha}\dot{\beta}}=(-i\sigma^2)_{\dot{\alpha}\dot{\beta}}$ (see e.g. \cite{Csaki2004}) we can adopt the following relations
\begin{equation}
	\chi_\alpha=\epsilon_{\alpha\beta}\chi^\beta\qc \psibar^{\dot{\alpha}}=\epsilon^{\dot{\alpha}\dot{\beta}}\psibar_{\dot{\beta}}\qc \qty\big(\chi^\dag)_\alpha=\chibar_{\dot{\alpha}}\qc \qty\big(\psibar^\dag)^{\dot{\alpha}}=\psi^{\alpha}.
\end{equation}
Therefore, the Dirac adjoint in terms of these spinors yields
\begin{equation}
	\bar{\Psi}=\left(\Psi_L^\dag,\Psi_R^\dag\right)\gamma^0=-\left(\qty\big(\psibar^\dag)^{\dot{\alpha}},\qty\big(\chi^\dag)_\alpha\right)=-\left(\psi^\alpha,\chibar_{\dot{\alpha}}\right).
\end{equation}
The symmetric Lorentz invariants from this notation are 
\begin{align}
	\chi\psi=\chi^\alpha\psi_\alpha &=\psi^\alpha\chi_\alpha=\psi\chi\\
	\chibar\psibar=\chibar_{\dot{\alpha}}\psibar^{\dot{\alpha}}&=\psibar_{\dot{\alpha}}\chibar^{\dot{\alpha}}=\psibar\chibar.
\end{align}


\section{Bessel equations and warped trigonometric functions}
\label{sec:appwarpedsinecosine}
\noindent
Starting with the equation of motion for the extradimensional profile of a left-handed 4D field
\begin{equation}
	\tLn''-\frac{4}{z}\tLn'+\qty\bigg(m_n^2-\frac{c_\psi^2+c_\psi-6}{z^2})\tLn=0\label{eq:appwarpedcasedecoupledEOM1},
\end{equation}
it is useful to first redefine $t_n\equiv z^{5/2}\tLntilde$. Plugging this into Eqs. \eqref{eq:appwarpedcasedecoupledEOM1} and dividing by $z^{1/2}$ yields
\begin{equation}
	z^2\tLntilde''+z\tLntilde'+\left(m_n^2z^2-\left(c_\psi+\frac{1}{2}\right)^2\right)\tLntilde =0\,,
\end{equation}
which corresponds to the Bessel equation 
\begin{equation}
	x^2\dv[2]{y}{x}+x\dv{y}{x}+(x^2-\nu^2)y=0,
\end{equation}
with $y\equiv \tLntilde$, $x\equiv m_nz$ and $\nu=1/2 + c_\psi$. For a right-handed field $\tilde{f}_n$ the same differential equation with $\nu=1/2-c_\chi$ is derived. Solutions to this differential equation will therefore be a combination of Bessel functions of the first and second kind
\begin{equation}
	y=A\BesselJ{\nu}{x}+B\BesselY{\nu}{x},
\end{equation}
with definitions
\begin{equation}
	\BesselJ{\nu}{x}=\sum_{k=0}^\infty\frac{(-1)^k\left(\frac{x}{2}\right)^{\nu+2k}}{k!\Gamma(\nu+k+1)}\quad\text{and}\quad\BesselY{\nu}{x}=\frac{\BesselJ{\nu}{x}\cos(\nu\pi)-\BesselJ{-\nu}{x}}{\sin(\nu\pi)},
\end{equation}
which means that the solutions of $\tLn$ and $f_n$ can be written as
\begin{align}
	\tLn(z)=z^{\frac{5}{2}}\qty\Big(\tilde{A}_n\BesselJ{\frac{1}{2}+c_\psi}{m_nz}+\tilde{B}_n\BesselY{\frac{1}{2}+c_\psi}{m_nz})\\
	f_n(z)=z^{\frac{5}{2}}\qty\Big(\tilde{C}_n\BesselJ{\frac{1}{2}-c_\psi}{m_nz}+\tilde{D}_n\BesselY{\frac{1}{2}-c_\psi}{m_nz}).
\end{align}
Since it is easier in some calculations throughout this work to deal with functions which possess some trigonometric properties, the mode basis will be redefined into
\begin{align}
	\tLn(z)&=\qty\bigg(\frac{R}{z})^{{c_\psi}-2}\qty\Big(A_nC_{c_\psi}(z)+B_nS_{c_\psi}(z))\\
		f_n(z)&=\qty\bigg(\frac{R}{z})^{-c_\psi-2}\qty\Big(C_nC_{-c_{\psi}}(z)+D_nS_{-c_{\psi}}(z)),
\end{align}
with
\begin{align}
	\label{eq:appwarpedcosinefunctionm}
	C_c(z)&\equiv \frac{\pi}{2}m_nR\qty\bigg(\frac{z}{R})^{c+\frac{1}{2}}\qty\Big(\BesselY{c-\frac{1}{2}}{m_nR}\BesselJ{c+\frac{1}{2}}{m_nz}-\BesselJ{c-\frac{1}{2}}{m_nR}\BesselY{c+\frac{1}{2}}{m_nz}) \\
	\label{eq:appwarpedsinefunctionm}
	S_c(z)&\equiv \frac{\pi}{2}m_nR\qty\bigg(\frac{z}{R})^{c+\frac{1}{2}}\qty\Big(\BesselJ{c+\frac{1}{2}}{m_nR}\BesselY{c+\frac{1}{2}}{m_nz}-\BesselY{c+\frac{1}{2}}{m_nR}\BesselJ{c+\frac{1}{2}}{m_nz})
\end{align}
and the relations
\begin{align}
	\tilde{A}_n&=m_nR^{-\frac{3}{2}}\frac{\pi}{2}\qty\bigg(A_n\BesselY{c-\frac{1}{2}}{m_nR}-B_n\BesselY{c+\frac{1}{2}}{m_nR})\\
	\tilde{B}_n&=m_nR^{-\frac{3}{2}}\frac{\pi}{2}\qty\bigg(B_n\BesselJ{c+\frac{1}{2}}{m_nR}-A_n\BesselJ{c-\frac{1}{2}}{m_nR})\\
	\tilde{C}_n&=m_nR^{-\frac{3}{2}}\frac{\pi}{2}\qty\bigg(C_n\BesselY{-c-\frac{1}{2}}{m_nR}-D_n\BesselY{-c+\frac{1}{2}}{m_nR})\\
	\tilde{D}_n&=m_nR^{-\frac{3}{2}}\frac{\pi}{2}\qty\bigg(D_n\BesselJ{-c+\frac{1}{2}}{m_nR}-C_n\BesselJ{-c-\frac{1}{2}}{m_nR}).
\end{align}
Using $J'_\nu(x)=\BesselJ{\nu-1}{x}-\nu/x\BesselJ{\nu}{x}$ and $Y'_\nu(x)=\BesselY{\nu-1}{x}-\nu/x\BesselY{\nu}{x}$ their derivative
\begin{align}
		C'_c(z)&= \frac{\pi}{2}m_n^2R\qty\bigg(\frac{z}{R})^{c+\frac{1}{2}}\qty\Big(\BesselY{c-\frac{1}{2}}{m_nR}\BesselJ{c-\frac{1}{2}}{m_nz}-\BesselJ{c-\frac{1}{2}}{m_nR}\BesselY{c-\frac{1}{2}}{m_nz}) \\
		S'_c(z)&= \frac{\pi}{2}m_n^2R\qty\bigg(\frac{z}{R})^{c+\frac{1}{2}}\qty\Big(\BesselJ{c+\frac{1}{2}}{m_nR}\BesselY{c-\frac{1}{2}}{m_nz}-\BesselY{c+\frac{1}{2}}{m_nR}\BesselJ{c-\frac{1}{2}}{m_nz})
\end{align}
can also be obtained quite easily. $S_C(z)$ and $C_c(z)$ are called warped sine and cosine functions because they show a similar behavior to the normal ones in the UV, i.e. $S_c(R)=0$, $C_c(R)=1$, $S_c'(R)=m_n$ and $C_c'(R)=0$.

It is more useful for calculations of the Higgs potential to work with the momentum instead of the mass. The expressions for these functions can be rewritten using $m_n=ip$. Doing this, one needs to make use of the modified Bessel functions which translate to the normal ones as
\begin{equation}
	\BesselI{\nu}{x}=i^{-\nu}\BesselJ{\nu}{ix}\quad\text{and}\quad\BesselK{\nu}{x}=\frac{\pi}{2}\frac{\BesselI{-\nu}{x}-\BesselI{\nu}{x}}{\sin(\nu\pi)}
\end{equation} 
with similar derivation rules  $I'_\nu(x)=\BesselI{\nu-1}{x}-\nu/x\BesselI{\nu}{x}$ and $K'_\nu(x)=-\BesselK{\nu-1}{x}-\nu/x\BesselK{\nu}{x}$. The warped trigonometric functions and their derivatives in terms of the momentum read
\begin{align}
	C_c(z)&=pR\qty\bigg(\frac{z}{R})^{c+\frac{1}{2}}\qty\Big(\BesselK{c-\frac{1}{2}}{pR}\BesselI{c+\frac{1}{2}}{pz}+\BesselI{c-\frac{1}{2}}{pR}\BesselK{c+\frac{1}{2}}{pz})\\
	S_c(z)&=ipR\qty\bigg(\frac{z}{R})^{c+\frac{1}{2}}\qty\Big(\BesselK{c+\frac{1}{2}}{pR}\BesselI{c+\frac{1}{2}}{pz}-\BesselI{c+\frac{1}{2}}{pR}\BesselK{c+\frac{1}{2}}{pz})\\
	C'_c(z)&=p^2R\qty\bigg(\frac{z}{R})^{c+\frac{1}{2}}\qty\Big(\BesselK{c-\frac{1}{2}}{pR}\BesselI{c-\frac{1}{2}}{pz}-\BesselI{c-\frac{1}{2}}{pR}\BesselK{c-\frac{1}{2}}{pz})\\
	S'_c(z)&=ip^2R\qty\bigg(\frac{z}{R})^{c+\frac{1}{2}}\qty\Big(\BesselK{c+\frac{1}{2}}{pR}\BesselI{c-\frac{1}{2}}{pz}+\BesselI{c+\frac{1}{2}}{pR}\BesselK{c-\frac{1}{2}}{pz}).
\end{align}
For small $m_nR\ll1$ and $m_nz\ll1$, the trigonometric functions can be approximated up to linear oder in $m_n$ as
\begin{equation}
    \label{eq:SCapproximations}
	C_c(z)=1+\order{m_n^2}\qquad S_c(z)=m_nR\frac{\left(\frac{z}{R}\right)^{2c+1}-1}{2c+1}+\order{m_n^3}.
\end{equation}


\section{Numerical calculation of the Higgs potential}
\label{sec:appHiggsPotential}
\noindent
Due to the quite lengthy expression of the spectral function for fermionic KK-towers (here for the top quark), it is beneficial for computational implementation of the derivatives of the Coleman-Weinberg potential to work in terms of the $10\times 10$ matrix $M_t$ and making use of trace techniques. For an easier calculation the derivations will be performed in terms of $s^2_h\equiv\sin^2(h/f)$ instead of $h$ with $s^2_{\tilde v}=\xi=v^2/f^2$. The potential and its derivatives, therefore, look like
\begin{align}
	\eval{V_t(h)}_{h=\tilde v}&=\eval{V_t(s_h^2)}_{s_h^2=\xi}=-\frac{3}{4\pi^2}\int_{0}^{\infty} \dd{p}p^3\log\qty\bigg(\frac{\det M_t(\xi)}{\det M_t(0)})\\
	\eval{\pdv{V_t(h)}{h}}_{h=\tilde v} &= \eval{\pdv{s_h^2}{h}}_{h=\tilde v}\eval{\pdv{V_t(s_h^2)}{s_h^2}}_{s_h^2=\xi}\notag\\
	&=-\frac{3\sqrt{\xi-\xi^2}}{2\pi^2f}\int_{0}^{\infty} \dd{p}p^3\trace\qty\bigg(M_t^{-1}(\xi)\eval{\pdv{M_t(s_h^2)}{s_h^2}}_{s_h^2=\xi})\\
	\eval{\pdv[2]{V_t(h)}{h}}_{h=\tilde v} &=\eval{\pdv[2]{s_h^2}{h}}_{h=\tilde v}\eval{\pdv{V_t(s_h^2)}{s_h^2}}_{s_h^2=\xi}+\eval{\qty\bigg(\pdv{s_h^2}{h})^2}_{h=\tilde v}\eval{\frac{\partial^2 V_t(s_h^2)}{(\partial s_h^2)^2}}_{s_h^2=\xi}\notag\\
	&=-\frac{3(1-2\xi)}{2\pi^2f^2}\int_{0}^{\infty} \dd{p}p^3\trace\qty\bigg(M_t^{-1}(\xi)\eval{\frac{\partial M_t(s_h^2)}{\partial s_h^2}}_{s_h^2=\xi})\notag\\
	&\phantom{=}-\frac{3\xi(1-\xi)}{\pi^2f^2}\int_{0}^{\infty} \dd{p}p^3\trace\bigg(M_t^{-1}(\xi)\eval{\frac{\partial^2 M_t(s_h^2)}{(\partial s_h^2)^2}}_{s_h^2=\xi}
	\notag\\
	&\phantom{=-\frac{3\xi(1-\xi)}{\pi^2f^2}\int_{0}^{\infty} \dd{p}p^3\trace\bigg(}-\qty\bigg[M_t^{-1}(\xi)\eval{\pdv{M_t(s_h^2)}{s_h^2}}_{s_h^2=\xi}]^2\bigg)= m_h^2.\label{eq:Higgsmassnum}
\end{align}


\bibliographystyle{JHEP}    
\bibliography{references}

\end{document}